\pdfoutput=1
\documentclass[letterpaper,twocolumn,10pt]{article}
\usepackage{booktabs}
\usepackage{tabularx}
\usepackage{amssymb}
\usepackage{usenix,epsfig}
\usepackage{grffile}
\usepackage[table,xcdraw]{xcolor}
 \usepackage{wrapfig, framed, caption}
\usepackage{subfigure}
\usepackage{xspace}
\usepackage[ruled, vlined, linesnumbered]{algorithm2e}
\usepackage[labelfont=bf]{caption}
\usepackage{tikz-qtree}

\usepackage{pifont}% http://ctan.org/pkg/pifont
\newcommand{\cmark}{\ding{51}}%
\newcommand{\xmark}{\ding{55}}%

\usepackage{hyperref}
\usepackage{xcolor}
\newcommand\crule[3][black]{\textcolor{#1}{\rule{#2}{#3}}}
\usepackage{enumitem,kantlipsum}
\usepackage[linewidth=0pt]{mdframed}
\mdfdefinestyle{myboxstyle1}{ backgroundcolor=gray!60}
\newboolean{draft}
\setboolean{draft}{true}
\ifthenelse{\boolean{draft}}{

}{}
\newcommand{\myfigure}[1]{Figure~#1}

\newcommand{\ie}{\textit{i}.\textit{e}., }
\newcommand{\eg}{\textit{e}.\textit{g}., }

\newcommand{\scout}{{\sc Scout}\xspace}

\usepackage{tikz}

\begin{document}

%don't want date printed
\date{}

%make title bold and 14 pt font (Latex default is non-bold, 16 pt)
% \title{\Large \bf Scout: Configuration Explorer in the Cloud}
\title{\Large Scout: An Experienced Guide to Find the Best Cloud Configuration}
% \title{An \textit{experienced} {\sc Scout} finds right cloud configurations}

%for single author (just remove % characters)
\author{
{\rm Chin-Jung Hsu},
{\rm Vivek Nair},
{\rm Tim Menzies},
{\rm Vincent W. Freeh}\\
North Carolina State University
% copy the following lines to add more authors
% \and
% {\rm Name}\\
%Name Institution
} % end author

\maketitle

% Use the following at camera-ready time to suppress page numbers.
% Comment it out when you first submit the paper for review.
\thispagestyle{empty}

\begin{abstract}

Finding the right cloud configuration for workloads is an essential step to ensure good performance and contain running costs.
A poor choice of cloud configuration decreases application performance and increases running cost significantly.
While Bayesian Optimization is effective and applicable to any workloads, it is fragile because performance and workload are hard to model (to predict).

In this paper, we propose a novel method, \scout. 
The central insight of \scout is that using prior measurements, even those for different workloads,
improves search performance and reduces search cost.
At its core,
\scout extracts search hints (inference of resource requirements) from low-level performance metrics.
Such hints enable \scout to navigate through the search space more efficiently---only spotlight region will be searched.

We evaluate \scout with 107 workloads on Apache Hadoop and Spark.
The experimental results demonstrate that our approach finds better cloud configurations with a lower search cost than state of the art methods.

Based on this work, we conclude that
(i) low-level performance information is necessary
for finding the right cloud configuration
in an effective, efficient and reliable way, and (ii)
a search method can be guided by
historical data, thereby reducing cost and improving performance.

\end{abstract}

\section{Introduction}
\label{sec:introduction}

Cloud computing provides a large variety of architectural configurations, such as
the number of cores, amount of memory, and the number of nodes.
The performance of a \emph{workload}---an application and its input---can execute up to 20 times longer---or cost 10 times more---than optimal.
The ready flexibility in cloud offerings has created a \emph{paradigm shift}.
Whereas before an application was tuned for a given cluster, in the cloud
the \emph{architectural configuration is tuned for the workload}.
Furthermore, because the cloud has a \emph{pay-as-you-go} model, each configuration (cluster size $\times$ VM type) has running cost and execution time.
Therefore, a workload can be optimized for least cost or shortest time---which are different configurations.

Choosing the right cloud configuration is a non-trivial problem for the following reasons.

\noindent\textbf{Brute Force.}
There are more than 100 cloud configurations. Consequently, evaluating all the possible configurations to find the best configuration is too expensive.

\noindent\textbf{Canonical Cloud Configuration.}
Each workload---application and input---has its own preferred choice of cloud configuration.
Therefore, there is no one-size-fits-all configuration~\cite{Alipourfard2017,Hsu2017,Venkataraman2016}. 

\noindent\textbf{Opaque Resource Requirements.}
Resource requirements to achieve a certain objective (execution time or running cost) for a specific workload are opaque~\cite{Yadwadkar2017}.

\noindent\textbf{Level playing field.}
While the execution time tends to decrease with a more powerful instance type, the cost per unit time goes up, which compresses the running costs. This creates a level playing field---several inferior configurations in execution time are now competitive in running cost~\cite{Hsu2017}.
Consequently, it is harder to find the optimal cloud configuration.

\iffalse
\begin{description}
\item[Brute Force:] There can be more than 100 cloud configurations. Consequently, evaluating all the possible cloud configurations to find the best configuration is too expensive.

\item[Canonical Cloud Configuration:]
Each workload---application and input---has its own preferred choice of cloud configuration.
Therefore, there is no one-size-fits-all configuration~\cite{Alipourfard2017,Hsu2017,Venkataraman2016}. 

\item[Opaque Resource Requirements:] Resource requirements to achieve a certain objective (execution time or running cost) for a specific workload are opaque~\cite{Yadwadkar2017}.

\item[Level playing field:] While the execution time tends to decrease with a more powerful virtual machine(VM), the cost per unit time goes up, which compresses the running costs. This creates a level playing field---several inferior configurations in execution time are now competitive in running cost~\cite{Hsu2017}.
More choices, much difficult.
\end{description}
\fi

Several methods have been proposed to find the best cloud configuration~\cite{Alipourfard2017, Venkataraman2016, Yadwadkar2017, Hsu2017}.
These methods can be broadly classified into 
(1) prediction---which uses elaborate offline evaluation to generate a machine learning model that predicts the performance of workloads and
(2) search-based techniques---which successively evaluate configurations looking for one that is near optimal~\cite{Alipourfard2017, Hsu2017}.
Prediction, as proposed in PARIS~\cite{Yadwadkar2017}, is not reliable because of high variance in prediction results. 
A search-based method does not require an accurate model but can have a high evaluation cost (measured in terms of configurations evaluated).

We choose the search-based method because it better tolerates prediction error and delivers effective solutions.
Any search-based method has two aspects.
\vspace{-0.5em}
\begin{itemize}[leftmargin=*]
    \setlength\itemsep{-0.4em}
    \item \textit{Exploration:} Gather more information about the search space by executing a new cloud configuration.
    \item \textit{Exploitation:} Choose the most promising configuration based on information collected.
\end{itemize}
\vspace{-0.5em}

Additional exploration incurs higher search cost, and insufficient exploration may lead to sub-optimal solutions.
This is the exploration-exploitation dilemma appeared in many machine learning problems~\cite{kaelbling1996reinforcement}.
For example, \emph{CherryPick} requires a good exploration strategy
to characterize the search space~\cite{Alipourfard2017}.

In this paper, we argue that it is possible to
trade exploration with exploitation without settling for a sub-optimal configuration.
The central insight of this paper is that the cost of the search for the right cloud configuration can be significantly reduced if we could learn from the \textit{historical data}---experiences of finding the right cloud configuration for other workloads.

In this paper, we present a \scout, which uses historical data to find the best cloud configuration for a workload. 
In doing so, we (1) enable practitioners to find a near-optimal cloud configuration (2) with a lower search cost than state of the art.
Additionally, we answer the following questions about improving the performance of the search-based method and reducing the search-cost.\\
\noindent\textbf{How to make the search process more \textit{efficient}.}
\emph{CherryPick} uses Bayesian Optimization to find the right cloud configuration~\cite{Alipourfard2017}.
It requires exploring the search space to characterize the performance of a workload on different cloud configurations.
Performance and workload characterization can be derived from data.
We show that using historical data (of other workloads),
\scout can largely reduce the cost of exploration.
Furthermore, \scout can find near-optimal configuration faster due to the experience learned from other workloads.\\
\noindent\textbf{How to make the search process more \textit{effective}.}
When a search process moves closer to the optimal solution at each step, it eventually finds the optimal configuration.
At its core, \scout predicts better choices (than the current best) at each step. 
\scout uses comprehensive performance data for the prediction.
\scout also leverages low-level performance information
to better identify resource requirements.
In this way, \scout can spotlight the region (such as cluster sizes and VM types) in the search space where near-optimal configurations reside.
Therefore, it is more likely to find effective solutions.\\
\noindent\textbf{How to make the search process more \textit{reliable}.}
In our previous work, we showed that \emph{CherryPick} is fragile because its search performance heavily depends on the selection of initial points (cloud configurations) and the choice of the kernel function~\cite{Hsu2017}.
More importantly, these choices are workload dependent.
This makes the search process unreliable or unpredictable.
We show that in \scout the search performance is more reliable regardless of initial points.

\noindent Our key contributions are:
\vspace{-0.5em}
\begin{enumerate}[leftmargin=*]
\setlength\itemsep{-0.4em}
\item we propose a novel method, \scout, that
finds (near) optimal solutions and solves the shortcomings of the prior work.
(Section~\ref{sec:approach});
\item we present a novel way to represent the search space, which can be used to transfer knowledge from historical measurements(Section~\ref{sec:approach});
\item we evaluate \scout and other state-of-the-art methods using more than 100 workloads on three different data processing systems.
(Section~\ref{sec:evaluation}); 
and
\item we make our performance data available for encouraging research of system performance.\footnote{
Large-scale performance data is available at \url{https://github.com/oxhead/scout.}}
\end{enumerate}
\vspace{-0.5em}
\section{Background and Related Work}
\label{sec:motivation}
\vspace{-0.4cm}

%Finding the best cloud configuration is essentially an optimization problem, which has been explored in many other fields of computer science~\cite{zhu2017bestconfig, herodotou2011starfish, bilal2017towards, Dalibard2017, nair2018finding, oh2017finding, Nair2017, Dewancker2015,shahriari2016taking,Klein2017,golovin2017google}
%Here, we only describe the most relevant work in finding the best cloud configuration.
The prior work in this area can be broadly divided into two: prediction and sequential model-based optimization.

\noindent\textbf{Prediction.} This technique builds a model to \emph{predict} the best architecture for a given workload.
Machine learning techniques have been widely used to build these prediction models.
Inside-out applies regression models to predict distributed storage performance (storage throughput)~\cite{Hsu2016}.
PARIS builds a complex performance model for batch-processing and OLAP jobs~\cite{Yadwadkar2017}.
Prediction accuracy heavily relies on feature selection, model selection, and parameter tuning and the quality of data.

\noindent\textbf{Sequential Model-Based Optimization.} A sequential model-based optimization (SMBO) method successively
finds towards the best cloud configuration.
During the search process, SMBO updates its ``belief'' (using new observations, running the workload on a cloud configuration) and builds a prediction model for selecting the next choice.
The prediction model built by SMBO is used to differentiate between suitable and unsuitable cloud configurations.
The search process terminates when the performance objective does not improve.
Bayesian Optimization falls into the SMBO class of algorithms.

\vspace{-0.3cm}
\subsection{Shortcomings of prior work}
\vspace{-0.15cm}
The state of the art techniques CherryPick~\cite{Alipourfard2017} and PARIS~\cite{Yadwadkar2017} suffer from three major issues. 

\noindent\textbf{Model accuracy} Prediction based approaches like PARIS build a model using measurements. The objective of such an approach is to use an accurate model to predict,
for example, execution time or running cost of workloads.
This method has two major weaknesses (i) building an accurate model requires more data---which in our setting is hard to come by, and (ii) the performance of the cloud environment is susceptible to performance variability---the data collected after running the workload might not reflect the true performance~\cite{tang2011impact}.
As shown in PARIS, the performance of batch-processing jobs is less predictable.
The inaccurate estimation of the execution time can be attributed to the non-linear relationship between resource and performance~\cite{Alipourfard2017}. 

\noindent\textbf{Cold-start} Any SMBO method requires initial measurements to seed the search process. The initial measurements are very crucial since it determines the effectiveness of the search. A poor seeding strategy can lead to wasted effort and leads to selecting a sub-optimal cloud configuration. The effect of cold-start is more pronounced when 
the initial measurement cost cannot be amortized,
\eg{the search space is not large enough.}

\noindent\textbf{Fragility} An SMBO method is fragile as it is overly sensitive to input parameters.
The success of \emph{CherryPick} on a given workload depends on the initial points used to seed the search and the choice of the kernel function used in the performance model (Gaussian Process Model).
In our previous work, we have observed that \emph{CherryPick}
sometimes fails to find near-optimal configurations and
incurs longer (than expected) search path~\cite{Hsu2017}.

\vspace{-0.3cm}
\subsection{Elements of an efficient approach}
\vspace{-0.15cm}
By analyzing the differences between the state of the art methods, we identified the
following key components in solving the problem: (1) a search-based method (similar to Cherrypick~\cite{Alipourfard2017}) is essential since it accommodates mispredictions and performance variances in the cloud, and (2) historical data (as used in PARIS~\cite{Yadwadkar2017}) is useful to understand the inherent preferences of a workload. 
These two components can be used together to solve the problem more effectively and overcome the shortcomings of the current state of the art approaches.

Because it is tough to build an accurate model
to directly predict performance and cost of workloads on distinct cloud configurations,
we can instead build an indirect model (for improving prediction accuracy).
A search-based method does not require a direct answer (which choice is the best), but an easier answer to ``are there better choices?''
That is, we can build a simplified and relaxed model that can assist a search-based method in finding the solutions more efficiently.
A \emph{relaxed} model does not predict the absolute performance of a configuration but rather predicts the relative performance of two configurations.
Furthermore, ``a bad learner'' sometimes can still find a good solution~\cite{Nair2017}.
Because SMBO requires an initial prediction model, it is necessary 
to \emph{pre-train}
However, the data for the initial model need not come from the workload being evaluated.
Rather, data from any workload can be used to build a useful model of the configuration space.
For example, \scout creates an initial model using data from more than 100 workloads across 69 configurations.
For this model to be most useful, the information must be generic---independent of workload. This technique is inspired by \cite{Yadwadkar2017}.\\
There are too few features (dimensions and options) in the configuration space to build a robust model that works across many workloads.
Consequently, a model based only on architectural features (\eg{cluster sizes and memory per core} is fragile.

Low-level performance metrics (which are generic) provide better insight into resource bottleneck and resource efficiency~\cite{Hsu2017, Hsu2016, Yadwadkar2017}.
Fortunately, this information is relatively easy and cheap to collect. 
\\
To summarize, the following elements are necessary
to create an effective approach.
\vspace{-0.5em}
\begin{enumerate}[leftmargin=*]
    \setlength\itemsep{-0.4em}
    \item Prefer the \textbf{\textit{search-based technique}}, which converges to the best solution iteratively and avoids the large penalty caused by dramatic prediction error.
    \item Use a \textbf{\textit{relaxed}} model that boosts prediction accuracy, thereby better guides a search process to find the near-optimal configurations more quickly,
    \item Use \textbf{\textit{low-level metrics}} to generate a generic representation of the search space such that it can be used by other combinations of workload and application.
    \item Create a \textbf{\textit{performance database}} so that the knowledge of optimization can be used by other optimizers to find the right cloud configuration and hence reduce the search cost. 
\end{enumerate}
\vspace{-0.5em}

\vspace{-0.4cm}
\section{From Observation to Action}
\label{sec:approach}
\vspace{-0.4cm}
In this section, we describe our method
of finding the right cloud configuration
for a given workload.
We first formalize our problem setting.
Next, we describe how to derive search hints from performance data
to guide a search process.
Last, we highlight the major design choices that enable an effective and efficient search method
for finding the best cloud configuration.

\vspace{-0.3cm}
\subsection{Problem Formulation}
\vspace{-0.15cm}
The search-based method attempts to find the best cloud configuration for a workload ($w\in W$)---an application and its input.
The cloud configuration space for workload $w$ is referred to as ($s\in S_w$), where $S_w$ is the set of cloud configuration options for a workload $w$. The size of the search space is $N_w$ cloud configurations. 
In our setting, the cloud configuration space is same for all workloads and hence referred to as $S$.
For a given workload $w$,
each configuration $s$ has a corresponding performance measure $y=\phi(s)$. Each configuration in the cloud configuration space $s$ is represented as the features of the cloud configuration. In CherryPick, the configuration is represented using architectural information such as the number of cores, size of memory, etc.

At the start of the search process for a workload ($w$), none of the cloud configurations are evaluated. The evaluated configurations are removed from a set of unevaluated configurations ($u\in U$) and added to a set of evaluation configurations ($E$). The search-based method evaluates different cloud configurations (follows a different path) to find the best configuration. Hence, the evaluated and unevaluated configuration for different workloads is varied. For every workload $w$, there exist a corresponding sets $U_w$ and $E_w$. The sum of the cardinalities of $U_w$ and $E_w$ is equal to the cardinality of $S$ ($|U_w| + |E_w| = |S|$). The performance data from other workloads ($H$) for a workload $w$ refers to the set of all evaluated cloud configuration of previous optimization process ($H_k = {E_x: \forall x=W\setminus k}$, where k is a workload).

A search method, while optimizing for a workload $w$, walks through the space of unevaluated configuration space $U_w$, and selects the next most promising configuration ($s_*\in U_w$). The promising configuration refers to the configuration, which has the highest probability of being better (lower execution time or deployment cost) than the configurations explored till until now ($P(s_*=\phi(s_*) < \phi(e) \forall e\in E_w)$).
To determine different choices for the next step, a search process only requires knowing ``how likely is one choice better than the others``.
\emph{CherryPick} chooses the regression model
for accurately predicting the actual performance of a workload.
Similarly, PARIS builds a regression model to estimate execution time and running cost on different VM types.

\vspace{-0.3cm}
\subsection{Searching for the Next Step}
\vspace{-0.15cm}
A search-based method navigates in the search space to find the best cloud configuration.
It is mostly concerned with
two questions:
``what are better choices'' and
``what are more promising regions''.
The former ensures that a search will
eventually, find a near-optimal configuration
while
the latter determines how quickly it finds the solution
(also known as convergence speed)
An effective and efficient search method must answer
these two questions.

To answer the two questions, we mainly need only to know ``what are better choices''. At each step, a search-based method aims to find a cloud configuration that is better than the current best.
A higher probability of \emph{guessing} the next step right
ensures that a search process sequentially finds a better choice.
A right next step also guides a search process to move towards the right direction.
As long as the optimizer can move closer to the desired solution at each step,
it is more likely to guarantee it will find
near-optimal solutions.

To better determine the next step, a search process can learn from the observations along the search path.
However, this method faces two challenges.
First, it requires collecting sufficient data to build strong beliefs.
\emph{Cherry\-Pick} is confronted by the cold-start issue since it must first ``explore'' the search space---to identify the promising regions---by building an accurate model.
Second, an insufficient number of observations leads to ``high bias'' in prediction---the method can wrongly believe that a particular region (such as VM types or cluster sizes) is more promising than the other leading to a sub-optimal solution.

Instead of learning only from observations collected while executing the workload, a search process also can learn from performance data of other workloads---which have been optimized in the past.
This addresses the issue of ``high bias'' because a larger number of measurements performance data is available to create a prediction model that generalizes a performance model better.
This also sidesteps the ``exploration'' problem because the search process does not need to collect observations by running workloads for the current search task. The idea of reusing the data is often tricky since the different combination of application and workload exhibit very different behavior.
For example, the same application with different inputs can create very different workload behavior (such as the execution time and running cost)~\cite{Hsu2017}.
A performance model, which captures this complex behavior would require more information about the search space than just the architecture level information such as VM types and cluster sizes.

\vspace{-0.3cm}
\subsection{Hints for Search}
\vspace{-0.15cm}
To navigate the search space efficiently, \scout is built on the following three ideas.

\noindent{\textbf{Relative ordering.}}
To effectively solve an optimization problem, we do not require accurate performance values rather; we care about relative ordering. This means that we do not need an accurate model as prescribed in prior work~\cite{Yadwadkar2017}. 
The idea of using an inaccurate model is useful because the effort required to build an accurate model is much higher than an inaccurate model. Hence, we do not need to predict performance measures directly rather the relative ordering or ranks.

\noindent{\textbf{Pairwise comparison.}}
Since the modeling scheme is only required to rank the set of unevaluated cloud configurations, we do not need to use regression-based modeling schemes as used in the prior work. Instead, we can use \textit{Pairwise Comparison} modeling scheme~\cite{wauthier2013efficient}.
Based on this insight, we choose not to infer (inaccurate) performance measure but rather to infer (accurate)
relative ordering---one configuration is better than another.
%a classification model which given a pair of configurations can infer is one configuration is better than other. 
Ranking from binary comparisons is a ubiquitous problem in many machine learning applications such as recommendation system and player ranking.

\noindent{\textbf{Transfer learning.}}
The accuracy of a performance model depends on the number of data points used to train the model. In our setting, the data points are expensive to compute. When building the performance models, it might be best to reference observations from other optimization processes. Researchers in transfer learning report that data from other optimization processes can yield better models than just using current data~\cite{peters2015lace2}.

Using these insights, \scout can built pairwise modelling technique to derive the probability $P(\phi(S_j) \leq \phi(S_i))$ where $S_j\in U_w$. Given any configuration pair $\langle S_i, S_j \rangle$, we derive the probability distribution over $U_w$.
A higher value indicates $S_j$ is more likely to have better performance measure than $S_i$. For example, consider a configuration space ($S={S_1, S_2, S_3, S_4}$) and $\phi(S_1) = 10$ (the current best), and the predicted probability distribution over $U_w = {S_2, S_3, S_4}$ (all pair-wise comparisons) is $[0.8, 0.2, 0.2]$. For a cloud configuration $S_i$, the vector of probability distribution over $U_w$ is referred to as $P_i$. This vector represents ``how likely $S_j$ ($S_j\in U_w$) is a better choice than $S_i$'',
where $S_i$ is the measured cloud configuration and $S_j$ is possible choices for the next step.
In this example, $S_2$ is a better choice than the others.
When the actual performance measure ($\phi(S_j) \leq \phi(S_i)$), a search process will find a better solution than the current best in the subsequent steps until termination.

\begin{figure}[t]
 \includegraphics[width=.4\textwidth]{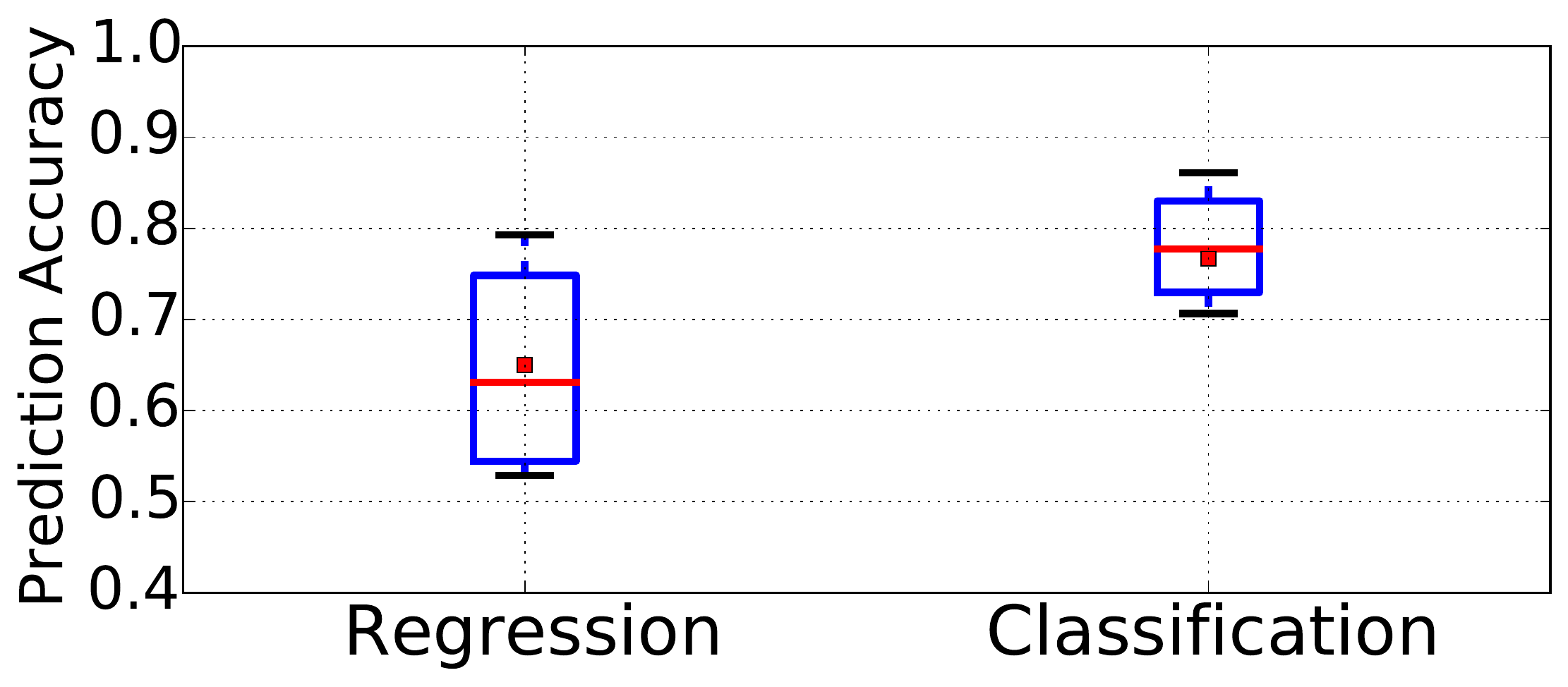}
 \centering
 \caption{\textbf{On the model selection of predicting the next step.}
 We evaluate the ability to distinguish a good and a bad configuration.
 In regression, we test rank preserving as prediction accuracy~\cite{Nair2017}.}
 \label{fig:reg_clas}
 \vspace*{-4mm}
\end{figure}

As mentioned above, we observe that a regression model performs worse than a pairwise model in terms of prediction accuracy. That is, the prediction of $\phi(S_j) \leq \phi(S_i)$ is less accurate when using a regression model. In our setting, accuracy quantifies how well a model can learn the probability of one configuration being better than the other configuration---which is similar to a classification task.
\myfigure{\ref{fig:reg_clas}} illustrates the difference in accuracy for all pairwise comparisons of configurations across all workloads using ExtraTrees~\cite{geurts2006extremely}.

Pairwise modeling is similar to a binary classification scheme. A classification problem is a problem of grouping data into classes. Traditional pairwise modeling is used as binary classification scheme.
But, in our setting, we use multi-classes instead of two.
The intuition for using multiple classes is that some configurations are very similar to each other.
In such a case, \scout should not select a configuration that produces a similar performance measure.
For example, we can define the classes to be ``better,'' ``fair,'' and ``worse''.
\scout favors the configurations in the ``better'' class.
\scout uses a predefined discretization policy (based on user-defined thresholds)  to convert probability to discrete classes.
For example, $\frac{\phi(S_j)}{\phi(S_i)} \leq 0.8$ is considered as ``better.''

Prior work used data from the current optimization process to explore the configuration space, which can be expensive. To reduce or eliminate the need for exploring, \scout uses historical data ($H_w$) gathered from previous runs. This is inspired by the field of Transfer learning, which focuses on learning a new
task through the transfer of knowledge from a related task that has already been explored. To transfer knowledge, \scout needs to encode the knowledge learned from the previous optimization in the historical data.
Our previous work showed that architectural features are not a reliable encoding scheme for defining the configuration space~\cite{Hsu2017}. Hence, %to solve the problem of transferring knowledge and improving search,
we use low-level metrics along with the architectural features.

\vspace{-0.3cm}
\subsection{Low-Level Insight}\label{sec:llm}
\vspace{-0.15cm}
The search performance of \scout is highly dependent on the classification accuracy.
%Performance of \scout is dependent on the classification accuracy of the performance model---high classification accuracy.
Low-level performance metrics helps identify
performance problems~\cite{Bodik2010, Novakovic2013} and
predict application and system performance~\cite{Hsu2016,Yadwadkar2017}.
Leveraging low-level information also alleviates the fragility problem of CherryPick.
In this section, we describe how to build an accurate classification model with low-level performance metrics.

Understanding resource bottlenecks help choose the right cloud configuration and help \scout ignore not so promising cloud configurations.
For example, in optimizing execution time, a memory bottleneck indicates an instance with larger memory may improves resource efficiency, thereby reducing execution time. However, resource requirement is opaque, and might not be linearly related to a performance objective~\cite{Alipourfard2017,Yadwadkar2017}.
In such cases, manual analysis is challenging and requires significant efforts.
Instead, we apply machine learning techniques to draw rules
from the historical data.

Let $L_i$ be the low-level metrics collected during the execution of a workload on
a cloud configuration $S_i$.
Our performance model learns the mapping function
$f(F(S_i), F(S_j), L_i) = {C_{ij}}$, where
the function $F$ represents the features of a cloud configuration, and
$C_{ij}$ is the prediction class that $S_{j}$ belongs to.
This prediction model answers the question ``provided low-level performance information to an observation of a workload running on $S_i$, will this workload perform better on another cloud configuration $S_j$?''
This is similar to the process of manual troubleshooting performance problems and identifying resource bottleneck. Instead of constructing rules manually, our modeling technique can extract those rules implicitly.
When a workload runs inefficiently on one cloud configuration,
\scout observes abnormal or insufficient resource usage.
This observation is translated to prediction probability implicitly.
\scout ignore those configurations with low prediction probability.

\vspace{-0.3cm}
\subsection{Search Strategy}
\vspace{-0.15cm}

During a search process, a new observation (running a workload on a selected cloud configuration) provides the necessary information to determine whether there exist other better choices. That is, given $\big \langle F(S_i), L_i \big \rangle$,
we can generate 
predict classes ($C_{ij}$).

The probability vector $P_{i}$ is derived for each new observation $\phi(S_{i})$.
A search strategy determines the choice based on the probability predictions.
At each step, the search process selects the configuration $S_j$ with
the maximum probability $P_{ij}$.

This search strategy is similar to depth-first search.
%\ick[---also exploitation first. -- what?]
While \emph{CherryPick} requires balancing exploration and exploitation,
\scout tends to exploit---since it uses historical data.
When the prediction model can generate quality predictions,
this search strategy leads to quick convergence speed
(the selected configuration improves over the current best).
Therefore, the search process requires low search cost of finding near-optimal configurations.

A search process should stop when it no longer can find a better configuration. This is controlled by a predefined parameter called  \textit{probability threshold} ($\alpha$) and acts as a stopping criterion.
When the predicted probability $P_{ij}$ is lower than $\alpha$ for all $S_j$,
the search process is not confident that it would find better configurations in the next step.
A search should also stop if it fails to find better solutions due to an inaccurate performance model. This is controlled by another parameter called \textit{ misprediction tolerance} ($\beta$) to avoid excessive search cost.

\vspace{-0.3cm}
\subsection{Put Them All Together}
\vspace{-0.15cm}

We have shown that the core element of a search based method is to determine the next best step. For obtaining hints to guide a search process, we propose using the classification technique (pair-wise prediction model) to predict the Probability of Improvement (PI). That is similar to Expected Improvement (EI) in \emph{Cherry\-Pick}.
We choose pair-wise modeling because it delivers high prediction accuracy and fits naturally into the search process.
The modeling method of \scout leverages low-level performance information, which extracts rules (based on resource utilization) implicitly.
This improves a search process because certain types of cloud configurations
can be avoided (as we will show in Section~\ref{sec:evaluation}).
Last, we choose a search strategy that merely picks the configuration with the highest prediction probability (most likely to be better than the current best).
This strategy increases convergence speed (low search cost).
\myfigure{\ref{fig:model_classification}} compares and contrasts the
design choices of \scout against prior work.

\begin{figure}
    \small{
        \begin{tabular}{@{}lcccc@{}}
        \toprule
        \textbf{Methods} & \textbf{\begin{tabular}[c]{@{}c@{}}Search-\\ based\end{tabular}} & \textbf{\begin{tabular}[c]{@{}c@{}}Low-level\\ Metrics\end{tabular}} & \textbf{\begin{tabular}[c]{@{}c@{}}Historical\\ Data\end{tabular}} & \textbf{\begin{tabular}[c]{@{}c@{}}Relaxed\\ Modeling\end{tabular}} \\ \midrule
        CherryPick~\cite{Alipourfard2017} & \cmark & \xmark & \xmark & \xmark \\
        PARIS~\cite{Yadwadkar2017} & \xmark & \cmark & \cmark & \xmark \\
        Arrow~\cite{Hsu2017} & \cmark & \cmark & \xmark & \xmark \\
        \cellcolor[HTML]{9B9B9B}\textcolor{white}{\textbf{Scout}} & \cellcolor[HTML]{9B9B9B}\color{white}\cmark & \cellcolor[HTML]{9B9B9B}\color{white}\cmark & \cellcolor[HTML]{9B9B9B}\color{white}\cmark & \cellcolor[HTML]{9B9B9B}\color{white}\cmark \\ %\bottomrule
        \end{tabular}
    }
    \centering
    \caption{
    \small{
    \textbf{An overall comparison with other methods in finding the best cloud configurations}.
    A search-based method better tolerates prediction bias.
    Leveraging low-level metrics improve search performance.
    Historical data helps eliminate unnecessary exploration overhead in a search.
    Pair-wise modeling naturally fits into a search-based method while improving prediction accuracy.}
    }
    \label{fig:model_classification}
\end{figure}

\vspace{-0.4cm}
\section{Evaluation}
\label{sec:evaluation}
\vspace{-0.4cm}

\begin{figure}[t]
\centering
\resizebox{0.9\linewidth}{!}{%
\small{
 \raggedleft
        \begin{tabular}{lrrrl}
        & \textbf{\{c4, m4, r4\}} & \textbf{\{c4, m4, r4\}} & \textbf{\{c4, m4, r4\}} & \\
        & \multicolumn{1}{c}{\textbf{large}} & \multicolumn{1}{c}{\textbf{xlarge}} & \multicolumn{1}{c}{\textbf{2xlarge}} & \\ \cline{2-2}
        %& \textbf{r4.large} & \textbf{r4.xlarge} & \textbf{r4.2xlarge} &  \\ \cline{2-2}
        \multicolumn{1}{l|}{} & \multicolumn{1}{r|}{4 $\times$ 2} & 4 $\times$ 4 & 4 $\times$ 8 &  \\ \cline{2-4}
        \multicolumn{1}{l|}{} & \multicolumn{1}{r|}{6 $\times$ 2} & 6 $\times$ 4 & \multicolumn{1}{r|}{6 $\times$ 8} & \textbf{Scale} \\ \cline{2-4}
        \multicolumn{1}{l|}{} & \multicolumn{1}{r|}{8 $\times$ 2} & 8 $\times$ 4 & 8 $\times$ 8 &  \textbf{Up}\\
        \multicolumn{1}{l|}{} & \multicolumn{1}{r|}{10 $\times$ 2} & 10 $\times$ 4 & 10 $\times$ 8 &  \\
        \multicolumn{1}{l|}{\textbf{Scale}} & \multicolumn{1}{r|}{12 $\times$ 2} & 12 $\times$ 4 & \cellcolor[HTML]{C0C0C0}12 $\times$ 8 &  \\
        \multicolumn{1}{l|}{\textbf{Out}} & \multicolumn{1}{r|}{16 $\times$ 2} & 16 $\times$ 4 &  &  \\
        \multicolumn{1}{l|}{} & \multicolumn{1}{r|}{20 $\times$ 2} & 20 $\times$ 4 &  &  \\
        \multicolumn{1}{l|}{} & \multicolumn{1}{r|}{24 $\times$ 2} & \cellcolor[HTML]{C0C0C0}24 $\times$ 4 &  &  \\
        \multicolumn{1}{l|}{} & \multicolumn{1}{r|}{32 $\times$ 2} &  &  &  \\
        \multicolumn{1}{l|}{} & \multicolumn{1}{r|}{40 $\times$ 2} &  &  &  \\
        \multicolumn{1}{l|}{} & \multicolumn{1}{r|}{\cellcolor[HTML]{C0C0C0}48 $\times$ 2} &  &  &  \\ \cline{2-2}
        \end{tabular}
}
}
 \caption{\small{\textbf{Search space that supports strong scaling.} We designed our search space such that we can evaluate the best cloud configuration to trade-off between the scale-out and the scale-up strategies. We are also interested in the best number of machines for a workload (\crule[gray!70]{0.23cm}{0.23cm} represent configurations with equal number of cores).
 We test 9 common instance types.
 }}
 \label{fig:search_space}
\end{figure}

We evaluate \scout with three sets of big data analytics
applications on 18 cloud configurations on a single-node.
We further evaluate 69 cloud configuration on multiple nodes.
Our evaluations show that \scout finds the optimal or near-optimal configuration more often than other methods
and does so while reducing search costs.

\vspace{-0.3cm}
\subsection{Experiment Setup}
\vspace{-0.15cm}
\label{sec:setup}

\noindent\textbf{Workloads.} We choose diverse workloads (CPU-intensive, memory-heavy, IO-intensive and network-intensive) such as PageRank, sorting, recommendation, classification and online analytical processing (OLAP).
We also change the input parameters and data sizes to create
a wide spectrum of workloads. 
These workloads run on Apache Hadoop~\cite{hadoop} and two separate versions of Apache Spark~\cite{Spark} (1.5 and 2.1). 
Please refer to \cite{Hsu2017} for detailed workload list.

\begin{figure}[ht]
 \centering
 \subfigure[Search Performance]{
 \label{fig:single_time_overall_performance}
 \includegraphics[width=.22\textwidth]{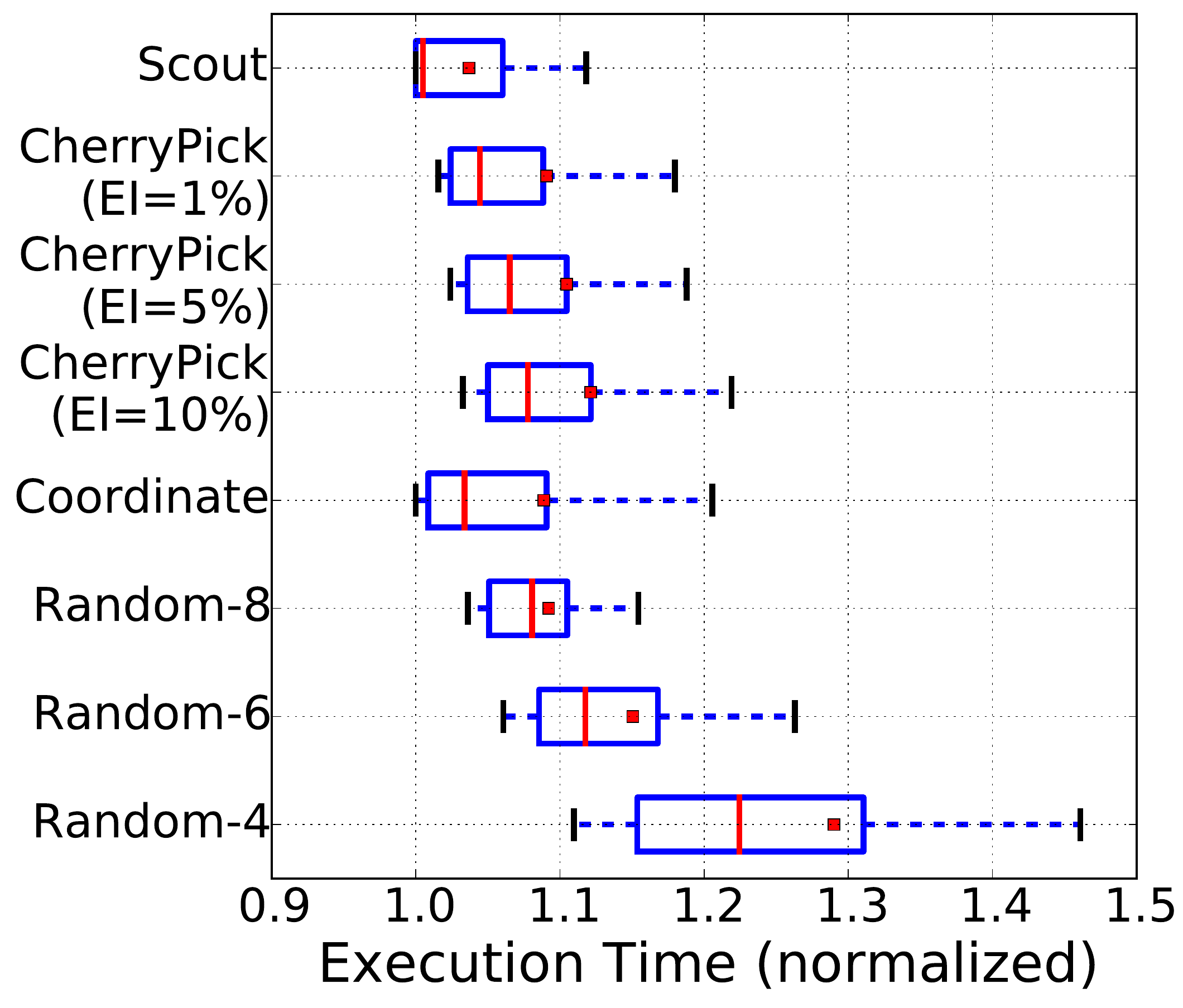}
 }%
 %\hspace{.1\textwidth}
 \subfigure[Search Cost]{
 \label{fig:single_time_overall_steps}
 \includegraphics[width=.22\textwidth]{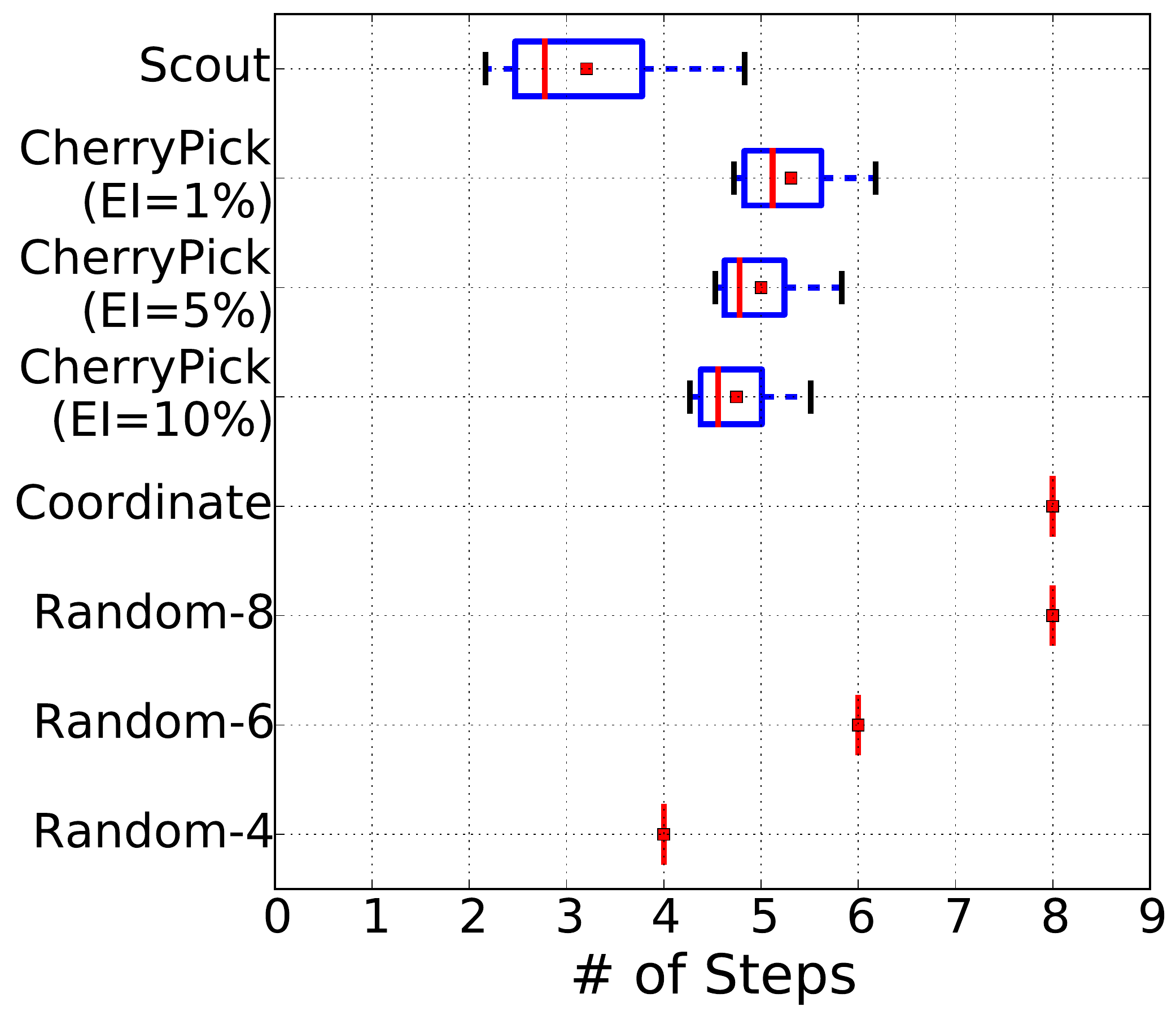}
 }
 \centering
 \caption{\small{\textbf{Minimizing Execution Time.}
 The \emph{x-axis} represents the normalized performance (to the optimal configuration), and the optimal performance is $1$. 
 \scout finds the near-optimal solutions ($< 1.1$) in 87\% workloads while using much fewer steps.}}
 \label{fig:single_time_overall}
 \vspace*{-4mm}
\end{figure}

\noindent\textbf{Deployment Choices.}
Our evaluation examines both single- and multiple-node settings.
When evaluating different cluster sizes,
we use strong scaling---fixed problem size---because we are interested in how to speed up a workload rather than the efficiency of the cluster.
The single-node setting serves a comparison baseline and allows us to test more workloads (due to smaller search space).
In the single node setting,
we choose 18 distinct instance types or cloud configurations and 107 workloads.
For the multiple-node setting, we run 18 workloads
on 69 cloud configurations (6 instances with various cluster sizes).
The search space is shown in~\myfigure{\ref{fig:search_space}}.

\noindent\textbf{Parameters.}
\scout has three important parameters:
1) labeled classes, 2) probability thresholds and 3) misprediction tolerance.
For the labeled classes in classification modeling,
we define five classes,
``better+'', ``better'',  ``fair'', ``worse'' and ``worse+'',
using thresholds [0.8, 0.95, 1.05, 1.2] as the cut points.
Regarding the two stopping criteria,
we choose $0.5$ for the probability threshold and
$3$ and $4$ for the misprediction tolerance
in the single-node and multiple-node setting respectively.
We examine the trade-off of these parameters in Section~\ref{sec:discussion}.

\noindent\textbf{Data Processing}:
The low-level performance data is collected using
the Linux performance monitoring tool, \emph{sysstat}~\cite{sysstat}.
We use a similar data processing method in our previous work~\cite{Hsu2016,Hsu2017}.

\vspace{-0.3cm}
\subsection{Comparison Method}
\vspace{-0.15cm}
To evaluate \scout, we examine the search performance
in terms of \emph{effectiveness}, \emph{efficiency} and \emph{reliability}.
We compare \scout with random search, coordinate descent, and \emph{CherryPick}.

\noindent\textbf{Random search.} This search method
uniformly samples the configuration space.
The stopping criterion is the number of
configurations to evaluate.
A higher number yields better solution but also incurs higher search cost.
For a fair comparison, the search is repeated 100 times.  Random-4, -6, -8 represent random samples of 4, 6, and 8 cloud configurations respectively.
It serves as a na\"ive baseline method.

\begin{figure}[ht]
 \centering
 \subfigure[Search Performance]{
 \label{fig:single_cost_overall_performance}
 \includegraphics[width=.22\textwidth]{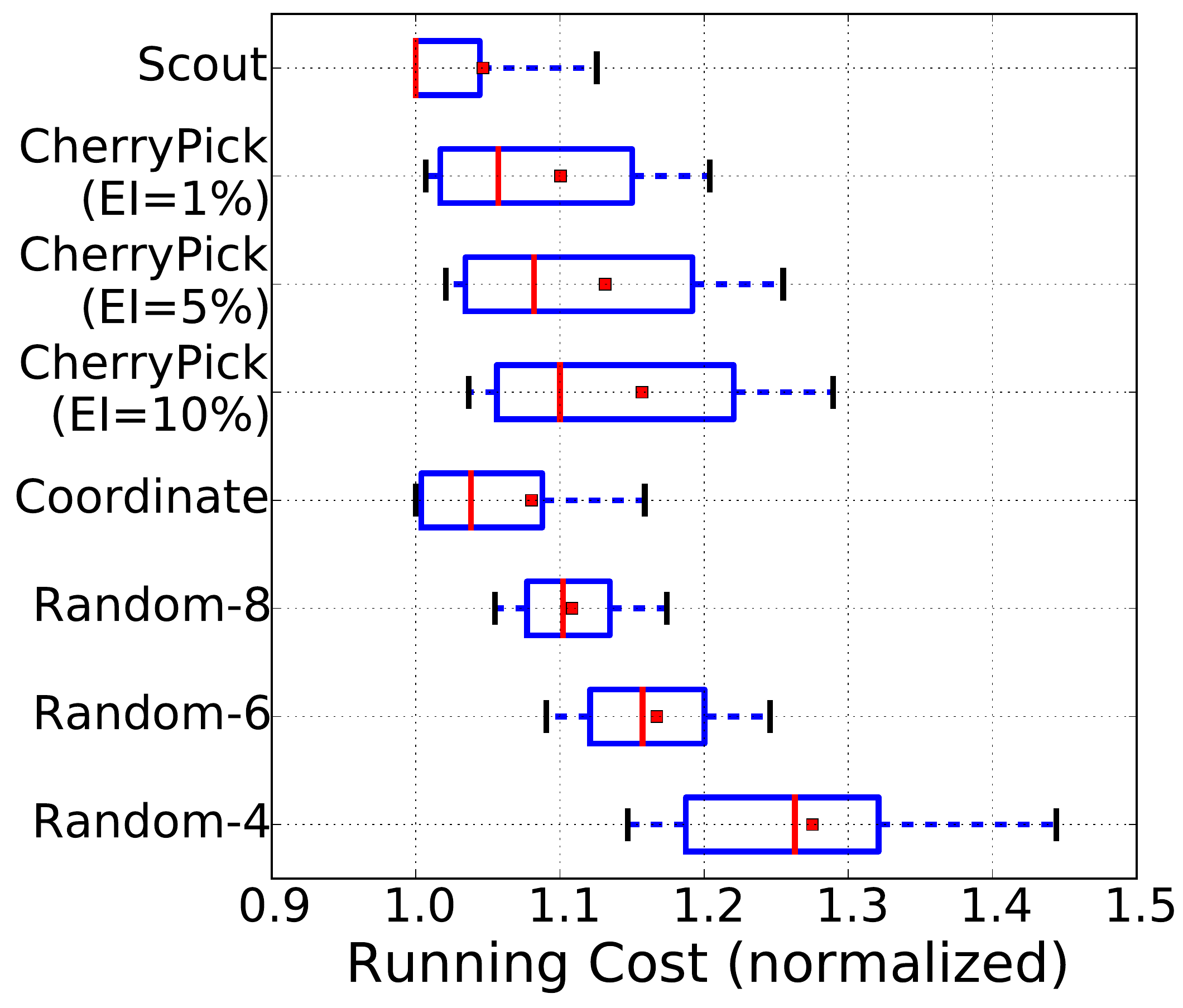}
 }%
 \subfigure[Search Cost]{
 \label{fig:single_cost_overall_steps}
 \includegraphics[width=.22\textwidth]{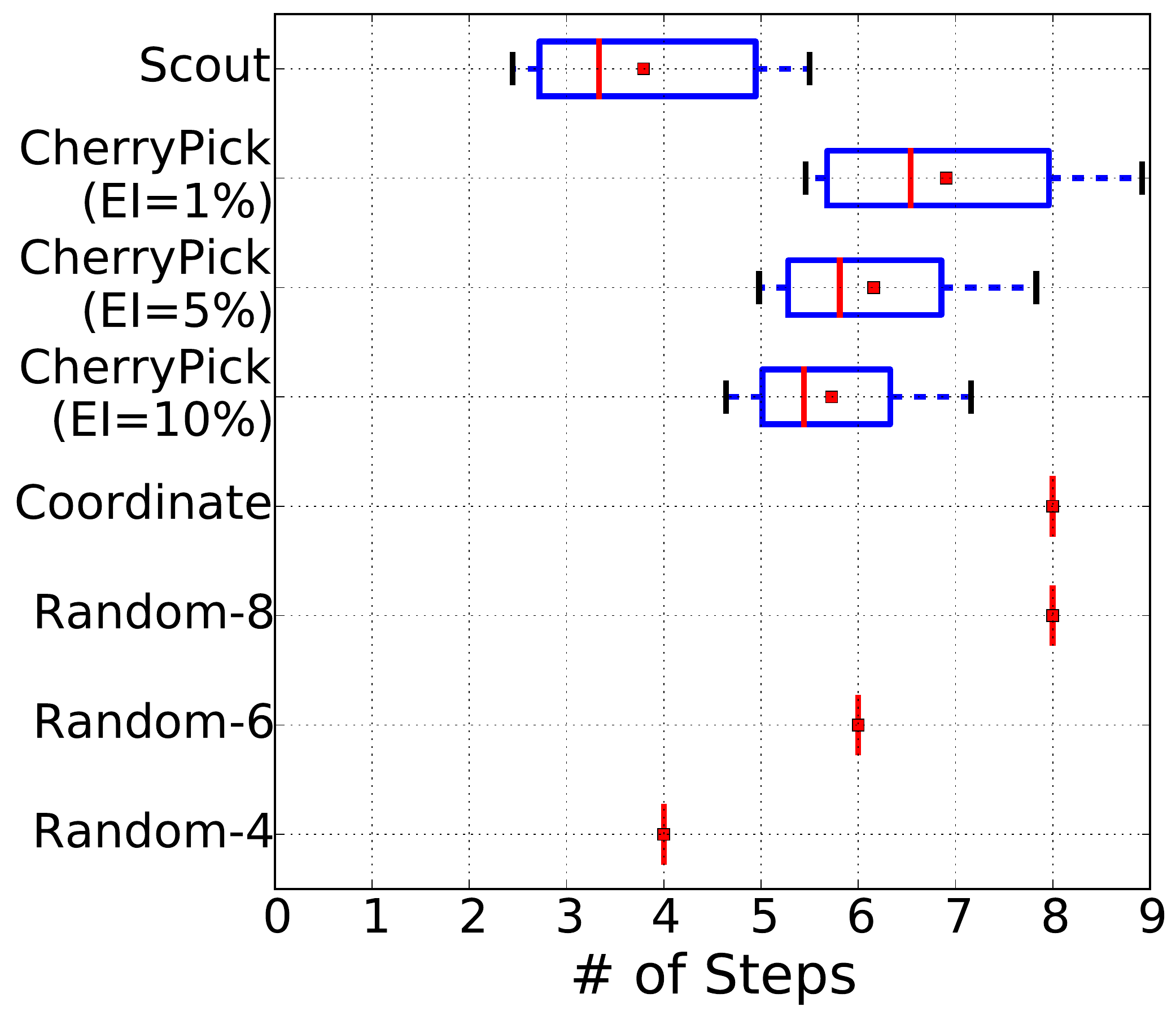}
 }
 \centering
 \caption{\small{\textbf{Minimizing Running Cost.}
 Searching for the optimal cost is more difficult because the search cost is higher than the scenario of minimizing execution time. \scout still finds near-optimal solutions with a small increase in search cost while \emph{CherryPick} only finds near-optimal solutions in about 50\% workloads.}}
 \label{fig:single_cost_overall}
 \vspace*{-4mm}
\end{figure}

\noindent\textbf{Coordinate descent.} This method searches
one dimension (\eg{CPU type and memory size}) at a time.
It determines the best choice of the dimension and
continues to choose the best from other dimensions.
This approach may suffer from local minimum due to
diminishing return and irregular performance outcome~\cite{Alipourfard2017}.
This situation worsens when the number of dimension increases.
In the evaluation, there are three dimensions:
(1) the instance family (such as \emph{c4} or \emph{r4}), (2)
the instance size (such as, \emph{large} or \emph{2xlarge}), and (3) the cluster size (\# of VMs).
The results are from 100 distinct searches, in which the starting point was randomly selected.

\noindent\textbf{CherryPick.} We implement the approach
proposed in \emph{CherryPick}~\cite{Alipourfard2017}.
We use
the same kernel function (\emph{Mat\'ern 5/2}) and
the same stopping criteria (\textbf{EI}=10\%).
We uniformly sample three configurations as starting points.
Since the search performance of \emph{CherryPick} is highly dependent on the selection of the starting points,
this experiment repeats 100 times to reduce artifacts
and give a better picture of \emph{CherryPick}'s capability.

\medskip
We compare these approaches using three metrics.
First, we evaluate the effectiveness of the methods using
the \textit{normalized performance} (to the optimal choice).
It can be the \textit{execution time} or the \emph{deployment cost}.
Second, we use the search cost--- the number of cloud configurations measured to find the right cloud configuration.
% We consider the number of steps instead of the search cost (the total cost of actual measurements) because the former reveals how fast a method finds a solution.
Last, we examine how reliable our method is across the workloads. % (in single node setting) and 18 (in multiple-node setting).
We compare the aggregate of the normalized performance and the search cost  along with their the 10\textsuperscript{th} and 90\textsuperscript{th} percentiles to observe whether our method performs well with consistency.  These numbers better illustrate reliability of the methods.

\begin{figure*}[t]
 \begin{minipage}[t]{0.32\linewidth}
    \includegraphics[width=5cm]{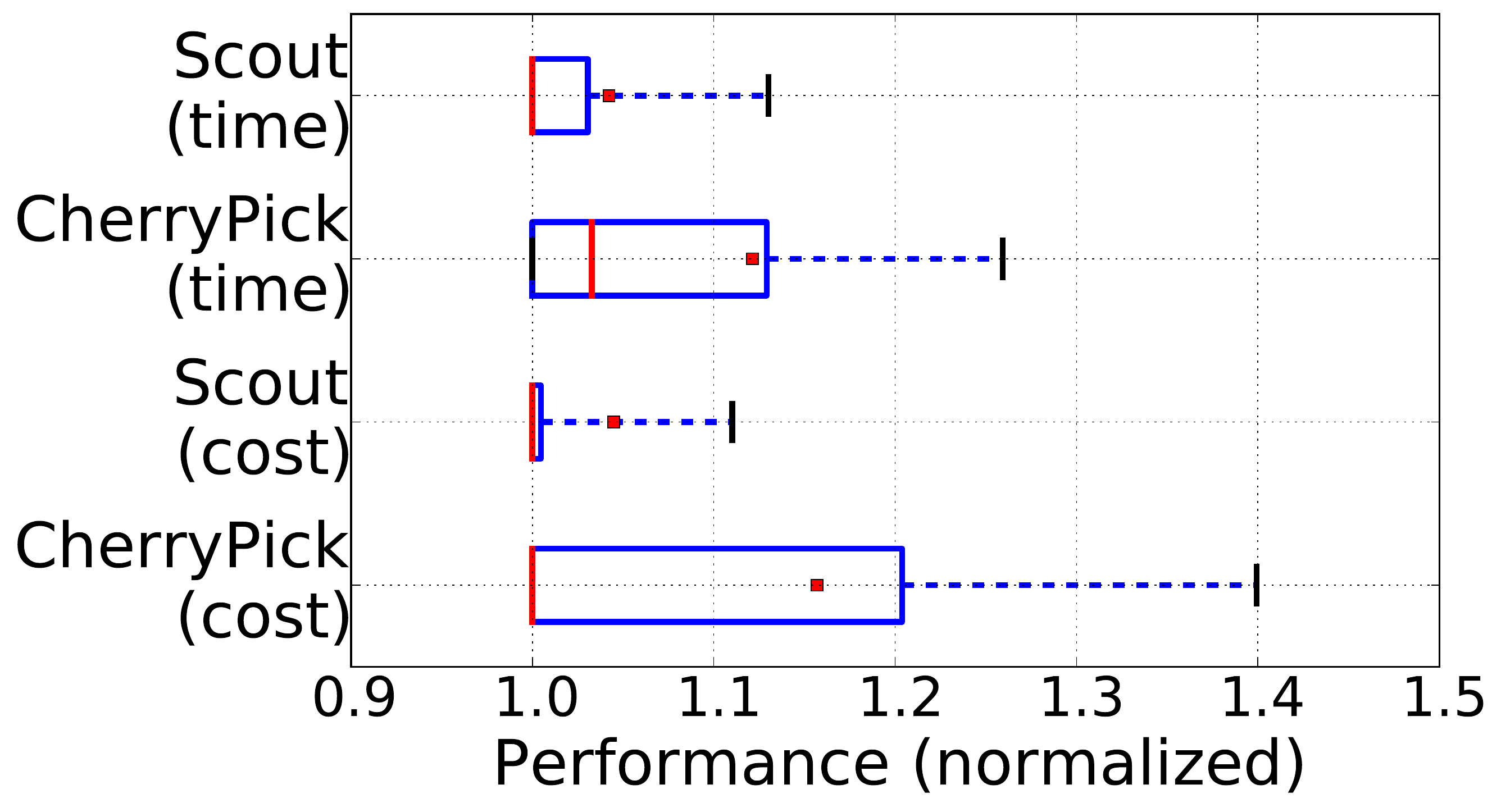}
    \centering
    \caption{\small{\textbf{Quality of found solutions.}
    Although both \emph{CherryPick} and \scout find the near optimal-solutions in most of the time,
    \scout is less fragile.}
    }
    \label{fig:single_fragility}
 \end{minipage}
 \hspace{0.2cm}
 \begin{minipage}[t]{0.32\linewidth}
    \includegraphics[width=5cm]{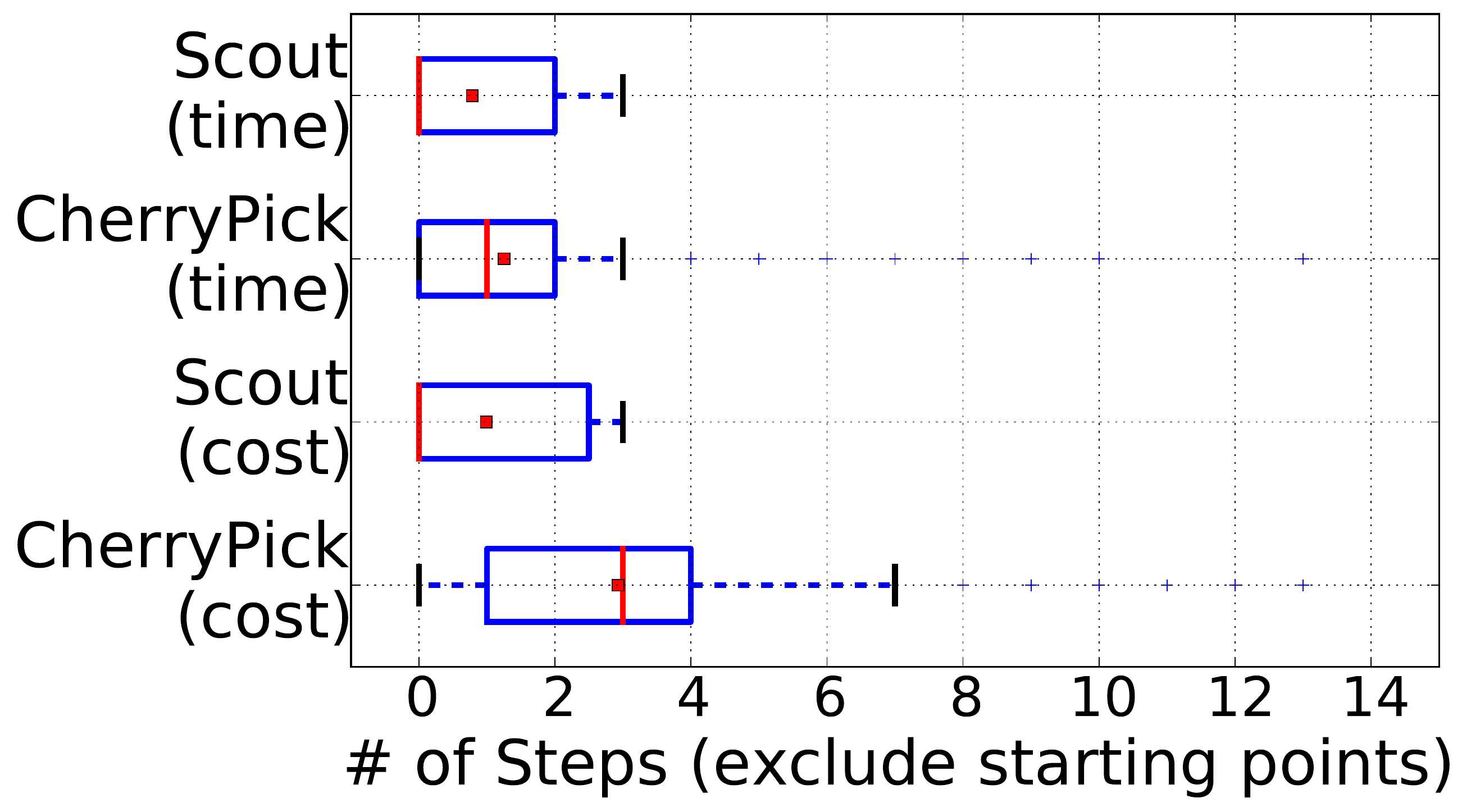}
    \centering
    \caption{\small{\textbf{Stopping awareness.}
    Search optimization avoids unnecessary search cost if it knows when the optimal solution is found.
    % \scout better acknowledges its existence.
    }}
    \label{fig:single_startingpoint}
 \end{minipage}
 \hspace{0.2cm}
 \begin{minipage}[t]{0.32\linewidth}
    \includegraphics[width=5cm]{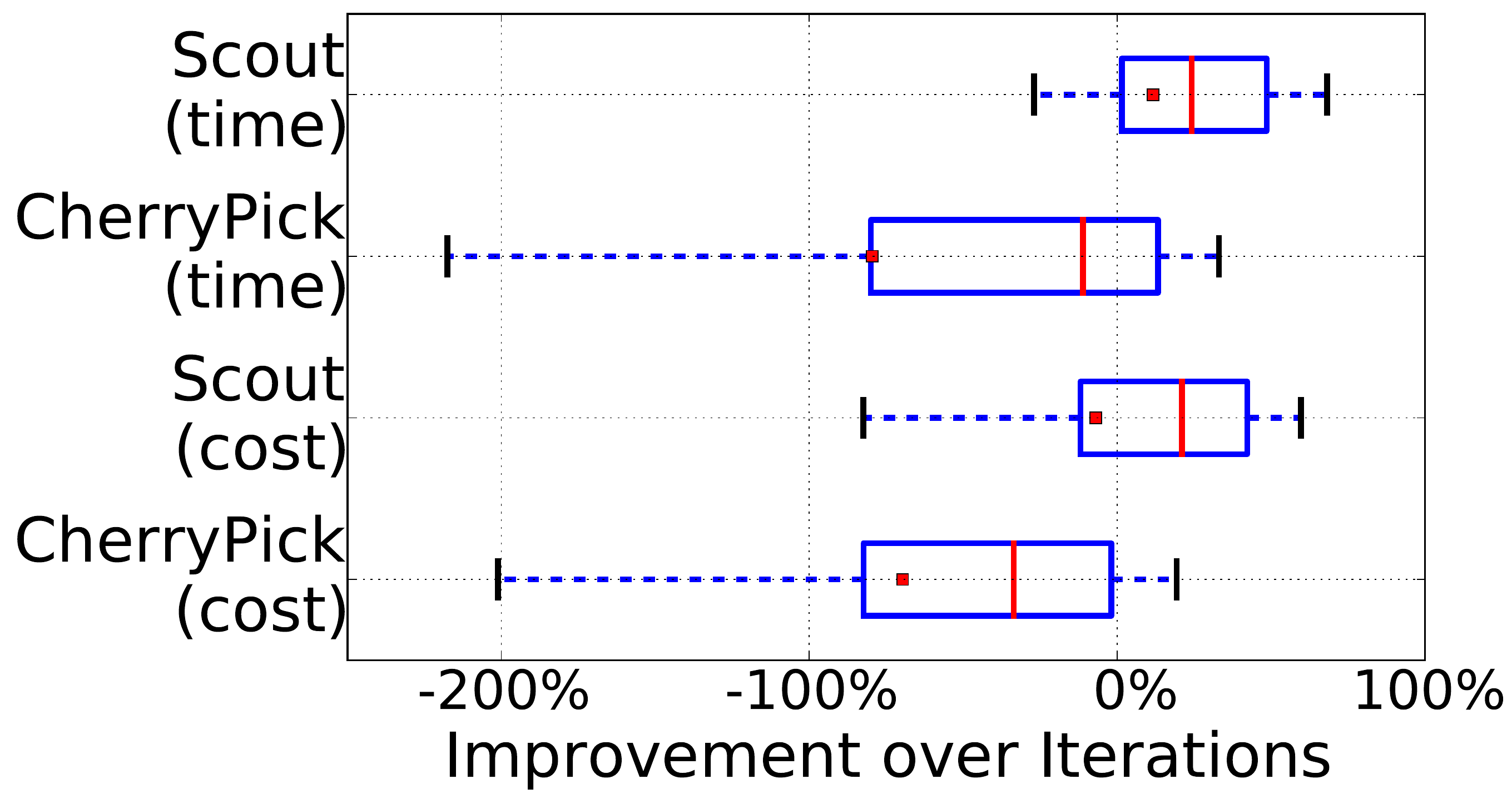}
    \centering
    \caption{\small{\textbf{Convergence speed.}
    \scout finds a better solution with 25\% improvement (on average) at each iteration, which suggests \scout is more likely to 
    converge.}}
    \label{fig:search_convergence}
 \end{minipage}
 \vspace*{-4mm}
\end{figure*}

\vspace{-0.3cm}
\subsection{Is Scout effective and efficient?}
\vspace{-0.15cm}

We examine search performance and search cost across 107 workloads in the single-node setting.
This evaluation largely answers whether a search method is reliable.

\textbf{Scout finds the near-optimal configurations
(within 10\% difference) for 87\% workloads}.
\myfigure{\ref{fig:single_time_overall_performance}} and \myfigure{\ref{fig:single_cost_overall_performance}} presents
the best cloud configuration (normalized to the optimal performance---1.0 represent the best, higher the worse) found by \scout and other methods while minimizing execution time and deployment cost, respectively. \myfigure{\ref{fig:single_time_overall_steps}} and \myfigure{\ref{fig:single_cost_overall_steps}} presents the search cost required the find the best cloud configuration which minimizes execution time and deployment cost, respectively. 
The figures display a box plot.
The box shows the inter-quartile range (from 25\textsuperscript{th} to 75\textsuperscript{th} percentile).
The vertical red line is the median, and the dot is the mean.
The whiskers to left and right show the 10\textsuperscript{th} and 90\textsuperscript{th} percentiles, respectively.
The horizontal axis shows execution time, and the vertical axis shows different techniques. 
An ideal search-based technique would find the best cloud configuration (in terms of performance) using the lower search cost.
These figures show the following.
\vspace{-0.5em}
\begin{itemize}[leftmargin=*]
    \setlength\itemsep{-0.4em}
    \item \scout finds the best relative performance in terms of both execution time and deployment cost. The median performance of \scout, while searching for the cloud configuration which minimizes the deployment cost is 1.0, which means \scout was able to find the right cloud configuration.
    \item \scout is better than CherryPick across all measures (execution time, deployment cost, and search cost).
    \item \scout finds the best relative performance using the least search cost (fewer number of steps). Random-4 also requires low search cost, but its performance is much worse than \scout.
    \item The variance in the performance (in terms of execution time and deployment cost) of \scout is much lower than the other methods. The large variance of the Random methods can be attributed to their inherent randomness.
\end{itemize}
\vspace{-0.5em}

Overall, we see that \scout is that best performing method and \emph{CherryPick}, the state of the art method, only delivers similar performance in 64\% workloads while requiring 47\% greater search cost (4.7 compared to 3.2 steps). We also observe that the variance in the best cloud configuration found by \scout over 100 runs across 107 workloads is much lower than the other method. Hence, we can conclude that \scout is a reliable method to find the best cloud configuration.

\noindent\textbf{Cost creates a level playing field.}
Optimizing execution time is relatively easy
because a larger, more powerful instance type is more likely to have a shorter execution time.
However, the more powerful types are more expensive to execute.
Consequently, a smaller instance type may run longer but cost less.
Because the cost to execute an instance grow as the raw hardware performance increase, the differences in deployment cost between configurations tend to be much less than the differences between execution time.
This levels the playing field for cost---many more configurations are good candidates.
This leveling leads to, in general, longer searches,
as shown in \myfigure{\ref{fig:single_cost_overall_steps}}.
\emph{CherryPick} requires one extra step in optimizing deployment cost with a 15\% decrease of workloads in which it fails to find a solution within 10\% of the optimal configuration. To summarize, the performance of a search-based method is dependent on the objective of the search. From the data, we observe that searching for the best cloud configuration in terms of cost is more challenging than finding the best cloud configurations in terms of execution time.

\vspace{-0.3cm}
\subsection{Is \scout reliable?}
\vspace{-0.15cm}
Users are willing to use a tool only when it is reliable. We evaluate the performance of \emph{CherryPick} and \scout
with different initial points for understanding their consistency.
In BO in \emph{CherryPick}, uses a random initial points to seed the search process and the effectiveness of CherryPick depend on these initial points. Selecting these initial points is non-trivial because (1) a good set of starting points for one workload does not work for other workloads, and (2) cloud providers frequently upgrade their instance portfolio with new instance types which make the process of selecting initial points more challenging. \scout is robust such that the effectiveness of \scout does not rely on initial points.

\begin{figure}[ht]
 \centering
 \subfigure[CherryPick]{
 \includegraphics[width=.22\textwidth]{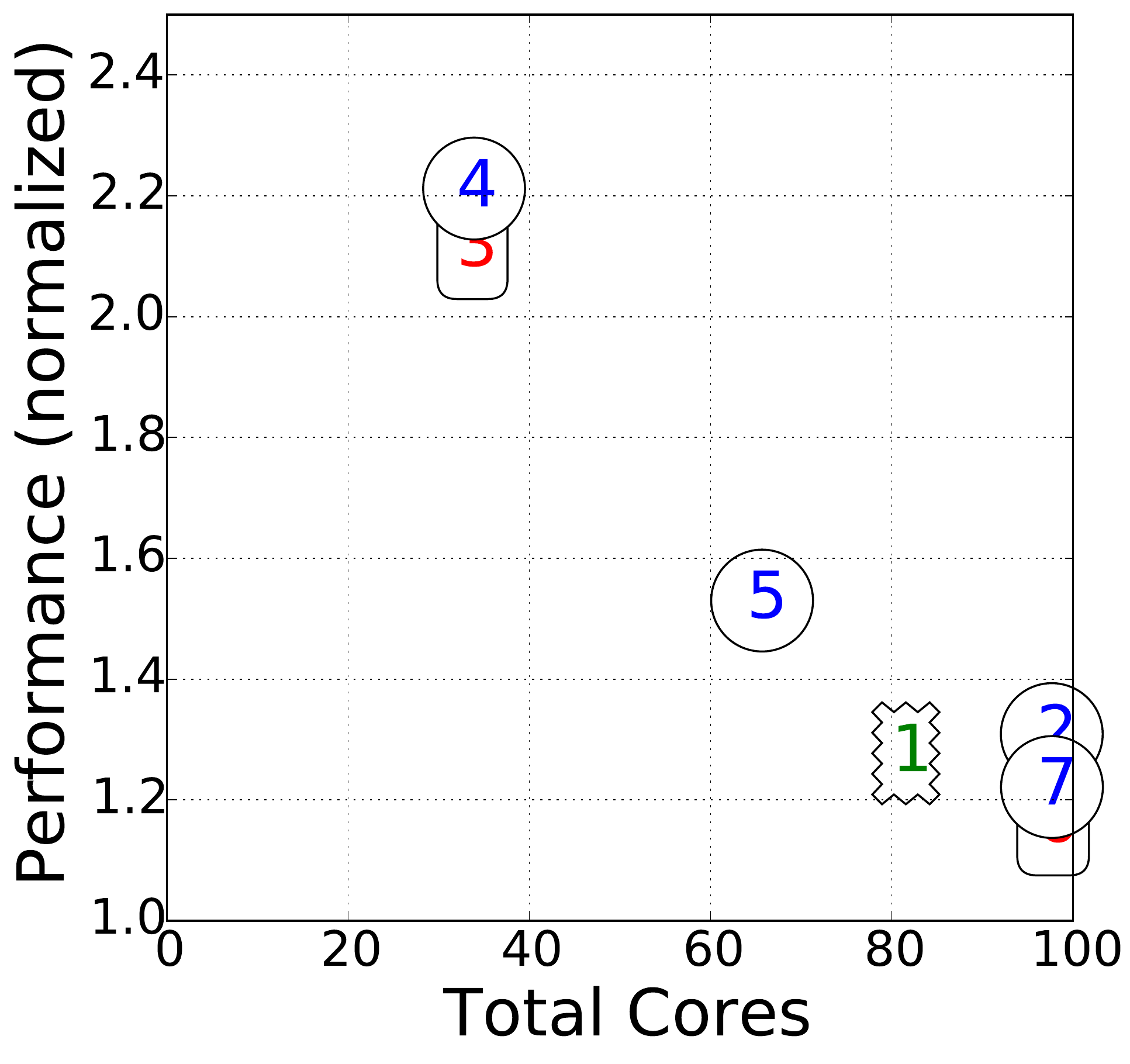}
 }%
 \subfigure[Scout]{
 \includegraphics[width=.22\textwidth]{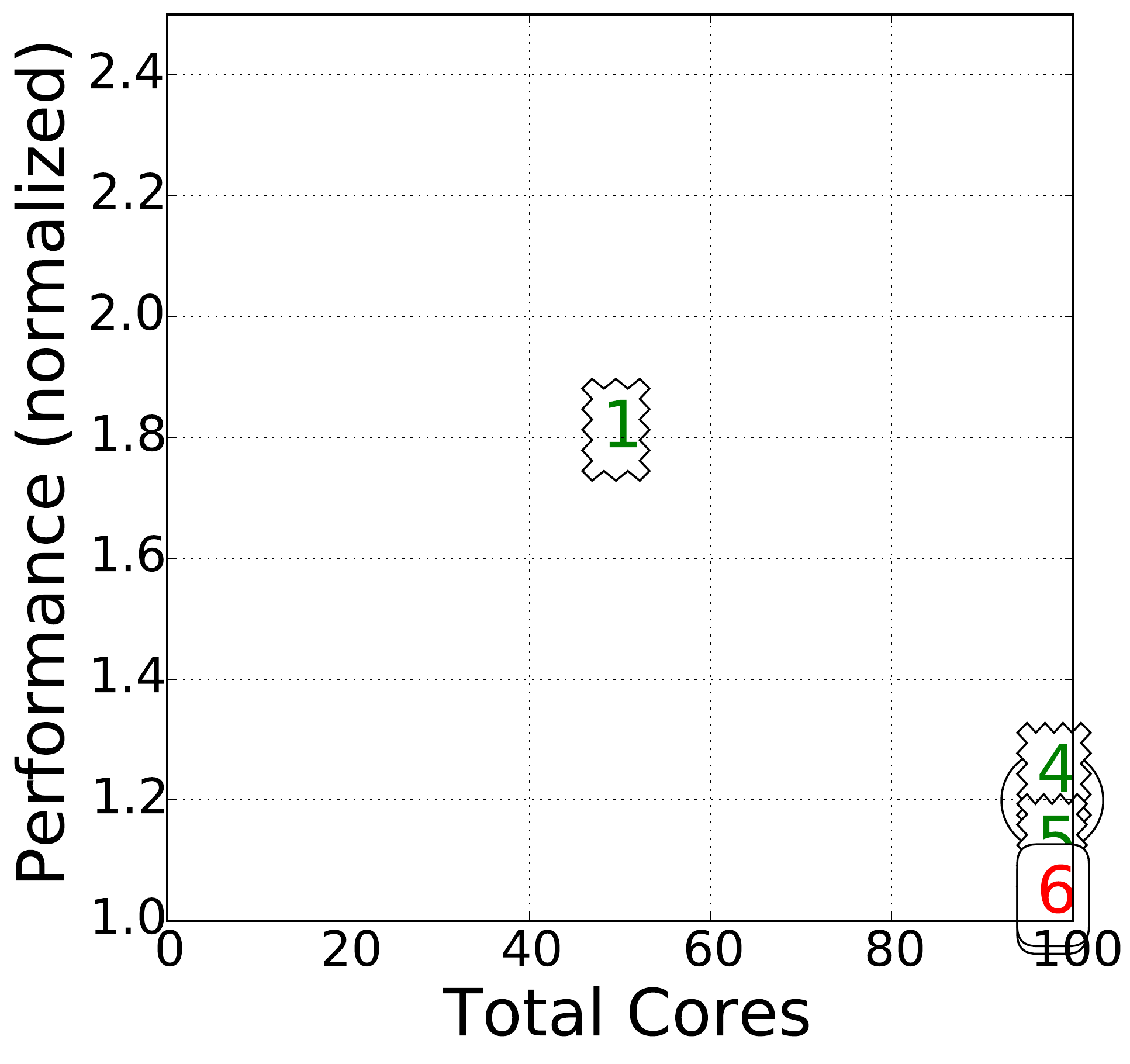}
 }
 \centering
 \caption{\small{\textbf{Finding the fastest configuration for PageRank on Hadoop.} Left \& right sub-figure show the search path of CherryPick and \scout respectively. \scout identifies PageRank as a compute-intensive workload.  It chooses the configurations with higher core counts and CPU speed.}}
 \label{fig:compare_1}
 \vspace*{-4mm}
\end{figure}

To demonstrate the robustness of \scout, for each workload, we varied the initial points used in \emph{CherryPick}.
These points were randomly (without replacement) selected from the search space. On the other hand, \scout only needs one starting point, which is also selected randomly. This experiment was repeated 100 times to understand the implication of randomness.
\myfigure{\ref{fig:single_fragility}} shows the variance in the normalized performance of the found solutions by both the methods.
We see that 
\vspace{-0.5em}
\begin{itemize}[leftmargin=*]
    \setlength\itemsep{-0.4em}
    \item \scout can find the optimal cloud configuration for most of the case since median performance is 1.0. However, there are some outliers which pushes the mean to 1.05. This is not a major concern since the 75$^{th}$ percentile is less than 1.05. This goes to show that the variance in the performance of 107 workloads aggregated over 100 runs is low.
    \item CherryPick is also effective in finding the cloud configuration since its median performance over 107 workloads is 1.05. We notice that the variance of the performance (both in terms of search performance and time) is larger than \scout. 
\end{itemize}
\vspace{-0.5em}

The variance in the results of CherryPick can be a major concern for the practitioners since a bad choice of initial points can lead to selecting either a slow or expensive configurations.
\scout, on the other hand, has more stable search performance regardless of the starting point.

\vspace{-0.3cm}
\subsection{Example Search Process}
\vspace{-0.15cm}

\begin{figure}[t]
 \centering
 \subfigure[CherryPick]{
 \includegraphics[width=.22\textwidth]{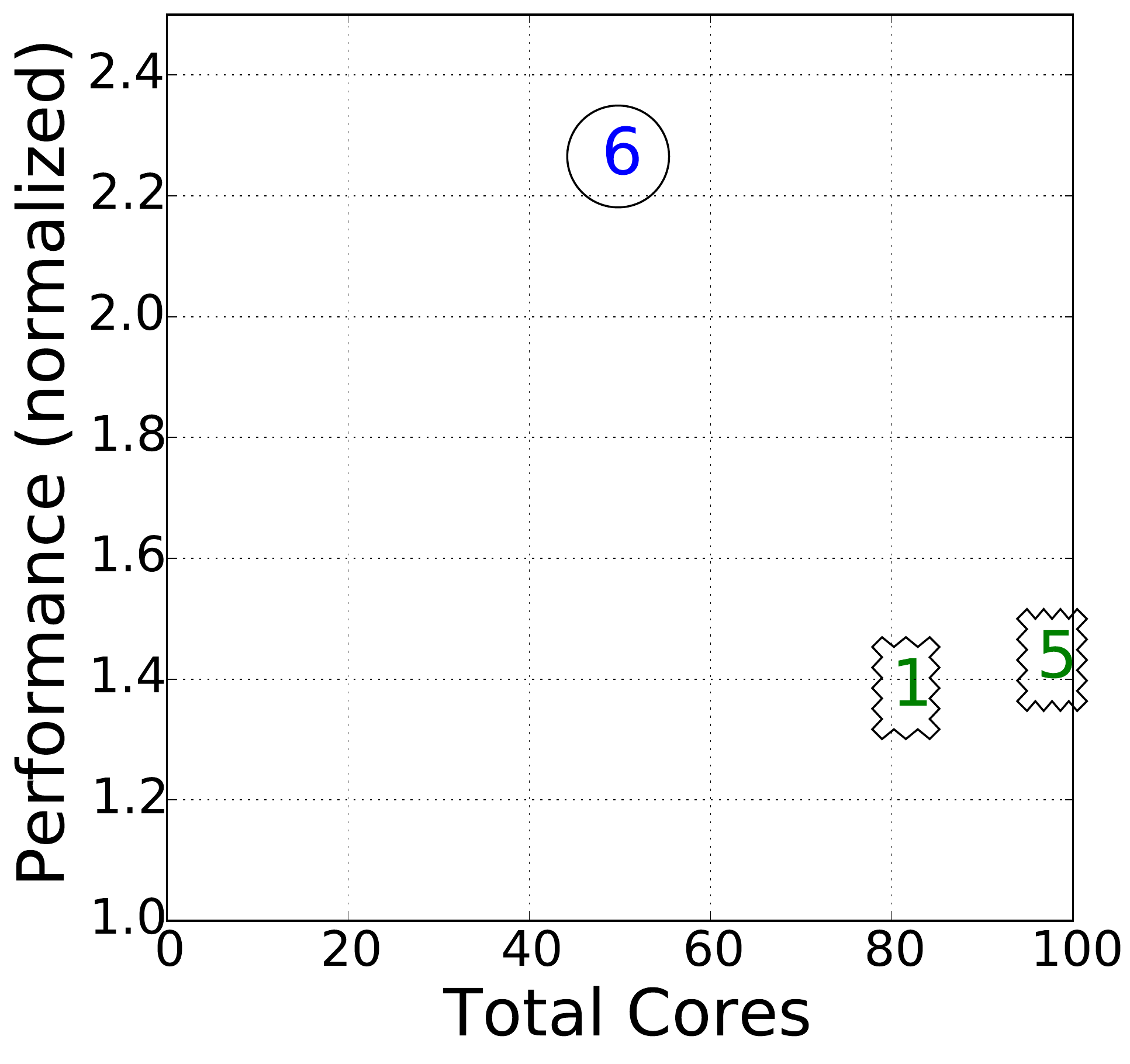}
 }%
 \subfigure[Scout]{
 \includegraphics[width=.22\textwidth]{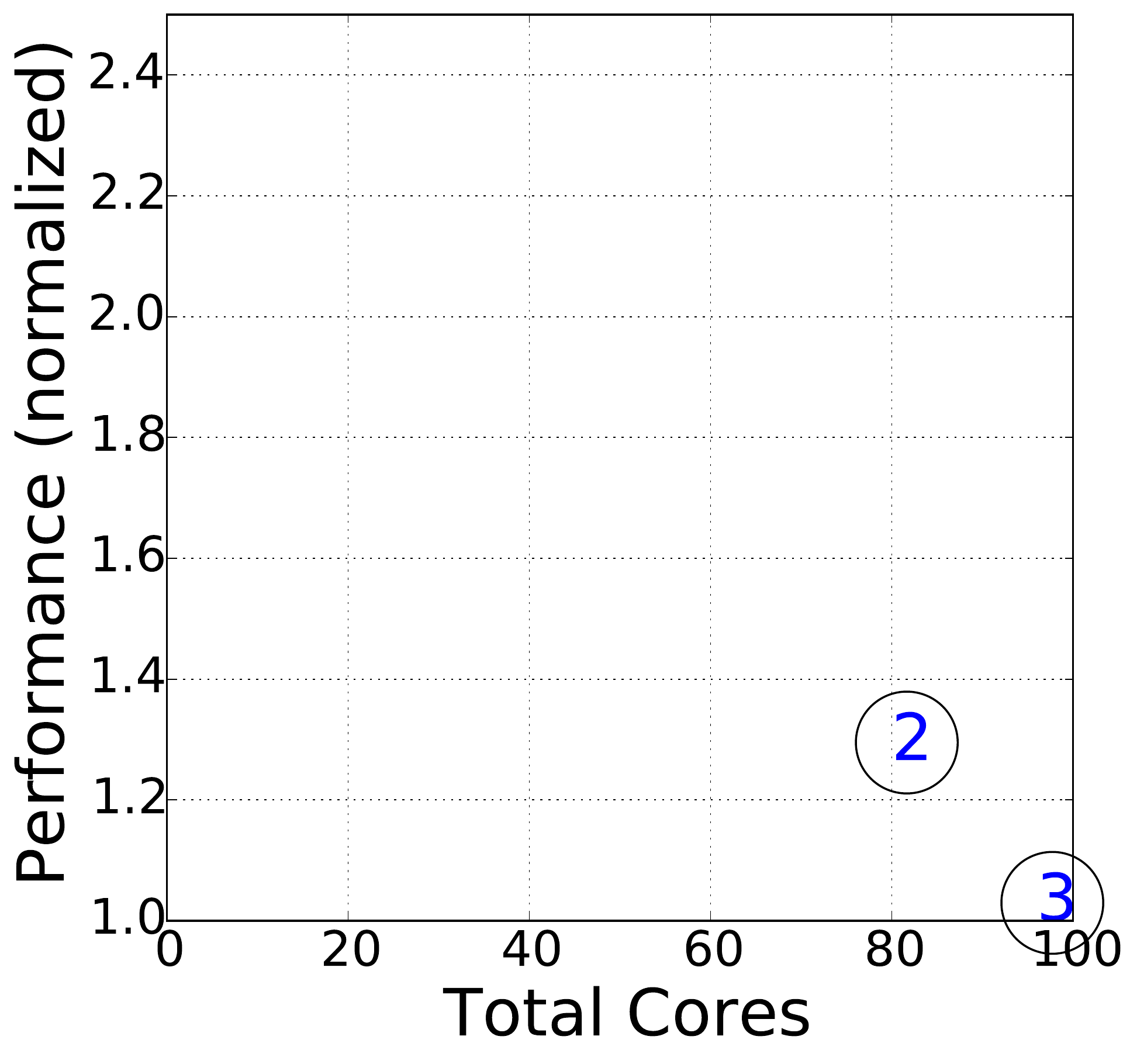}
 }
 \centering
 \caption{\small{\textbf{Minimizing execution time of Regression on Spark.} Since the Regression workload requires both computation and large memory, \scout directly chooses configurations with the \emph{r4} family and larger cores.}}
 \label{fig:compare_2}
 \vspace*{-4mm}
\end{figure}

\begin{figure}[ht]
 \centering
 \subfigure[CherryPick]{
 \includegraphics[width=.22\textwidth]{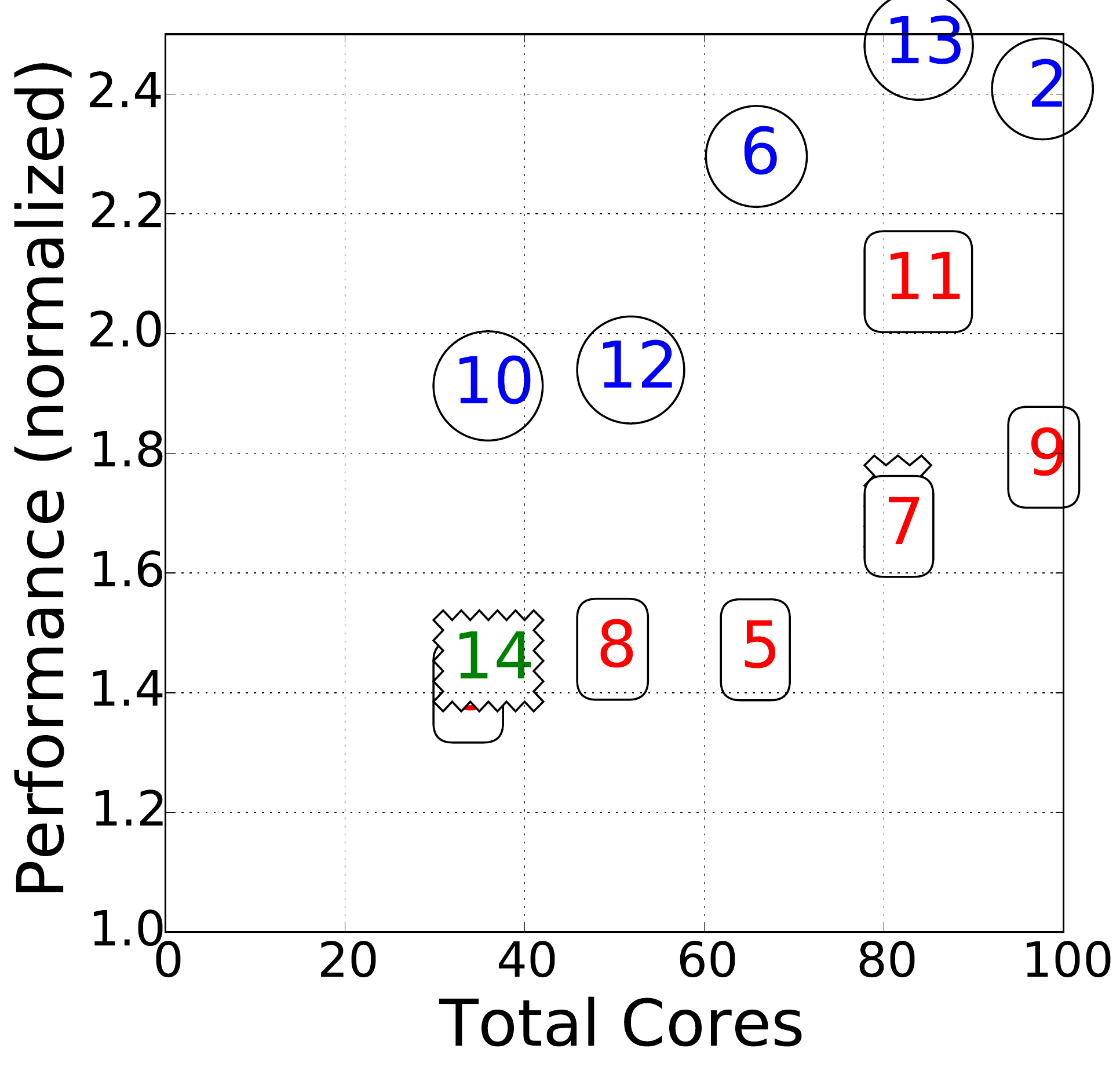}
 }%
 \subfigure[Scout]{
 \includegraphics[width=.22\textwidth]{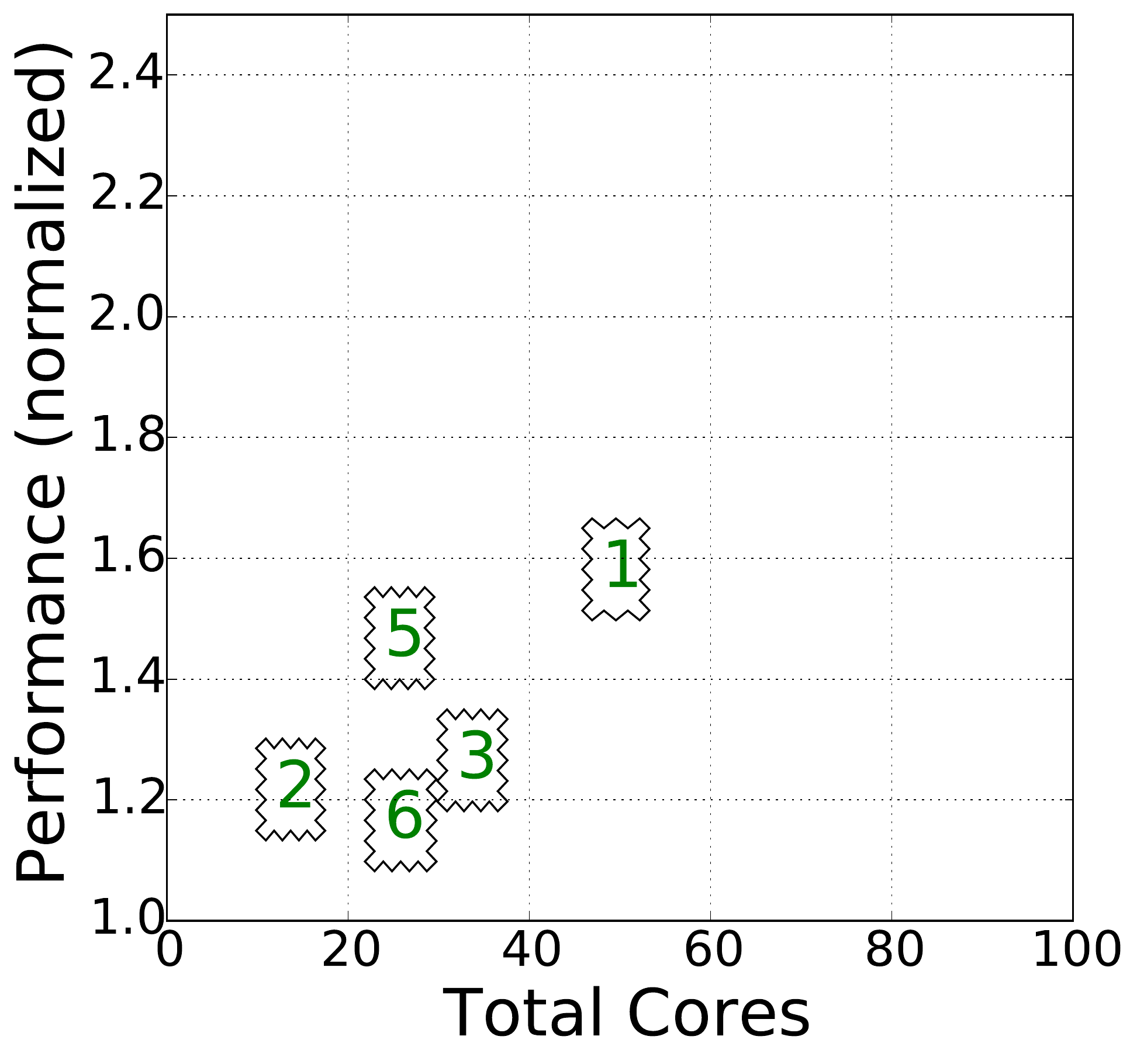}
 }
 \centering
 \caption{\small{\textbf{Finding the cheapest configuration for Terasort on Hadoop.} The Terasort workload requires enough memory to avoid spilling data to disks. Besides, a large cluster can be insufficient due to the shuffle phase in MapReduce.  \scout chooses a smaller cluster with the general-purpose VM type. }}
 \label{fig:compare_3}
 \vspace*{-4mm}
\end{figure}

This section compares and contrasts the properties of \emph{CherryPick} and \scout.
We provides four examples of optimizing execution time (in Figure~\ref{fig:compare_1} and~\ref{fig:compare_2}) and running cost (in Figure~\ref{fig:compare_3} and~\ref{fig:compare_1}). Different colored markers in the graphs represent different families of instances: \textcolor{green}{green} represent the m4 family---general purpose, \textcolor{blue}{blue} represent the r4 family---memory optimized, and \textcolor{red}{red} represent the c4 family---compute optimized.
We evaluate \emph{CherryPick} and \scout on four representative workloads, selected based on diverse resource requirements (CPU intensive, Memory intensive).
For \emph{CherryPick}, we choose
\emph{20$\times$m4.xlarge}, \emph{48$\times$r4.large} and \emph{16$\times$c4.large} as the starting points because
they are wide spread in the search space.
Since \scout only needs one starting point,
we choose \emph{24$\times$m4.large} because it is the mid point of the search space.
We observe that \emph{CherryPick} can find near-optimal solutions for few workloads if not all.

\begin{figure}[ht]
 \centering
 \subfigure[CherryPick]{
 \includegraphics[width=.22\textwidth]{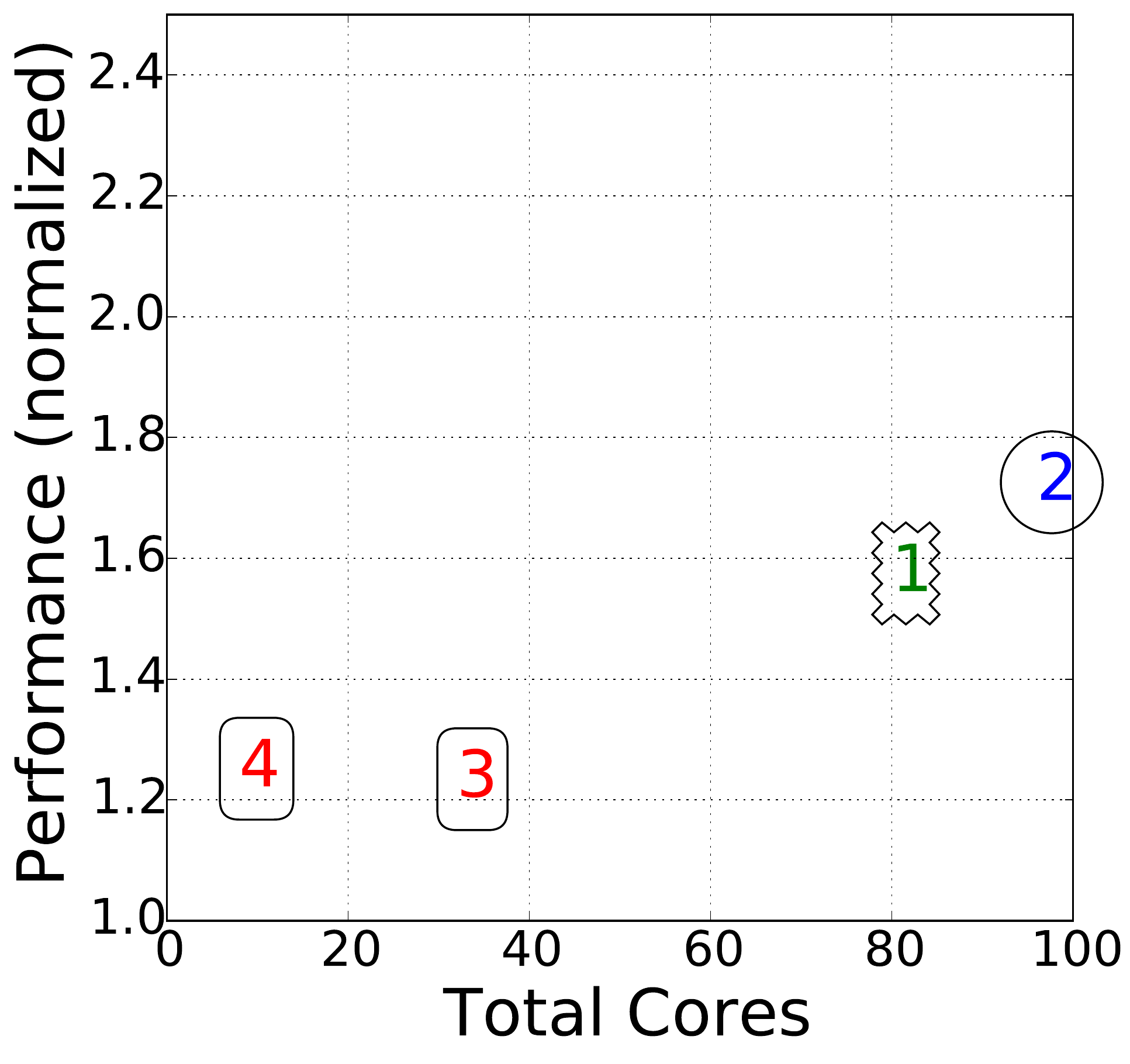}
 }%
 \subfigure[Scout]{
 \includegraphics[width=.22\textwidth]{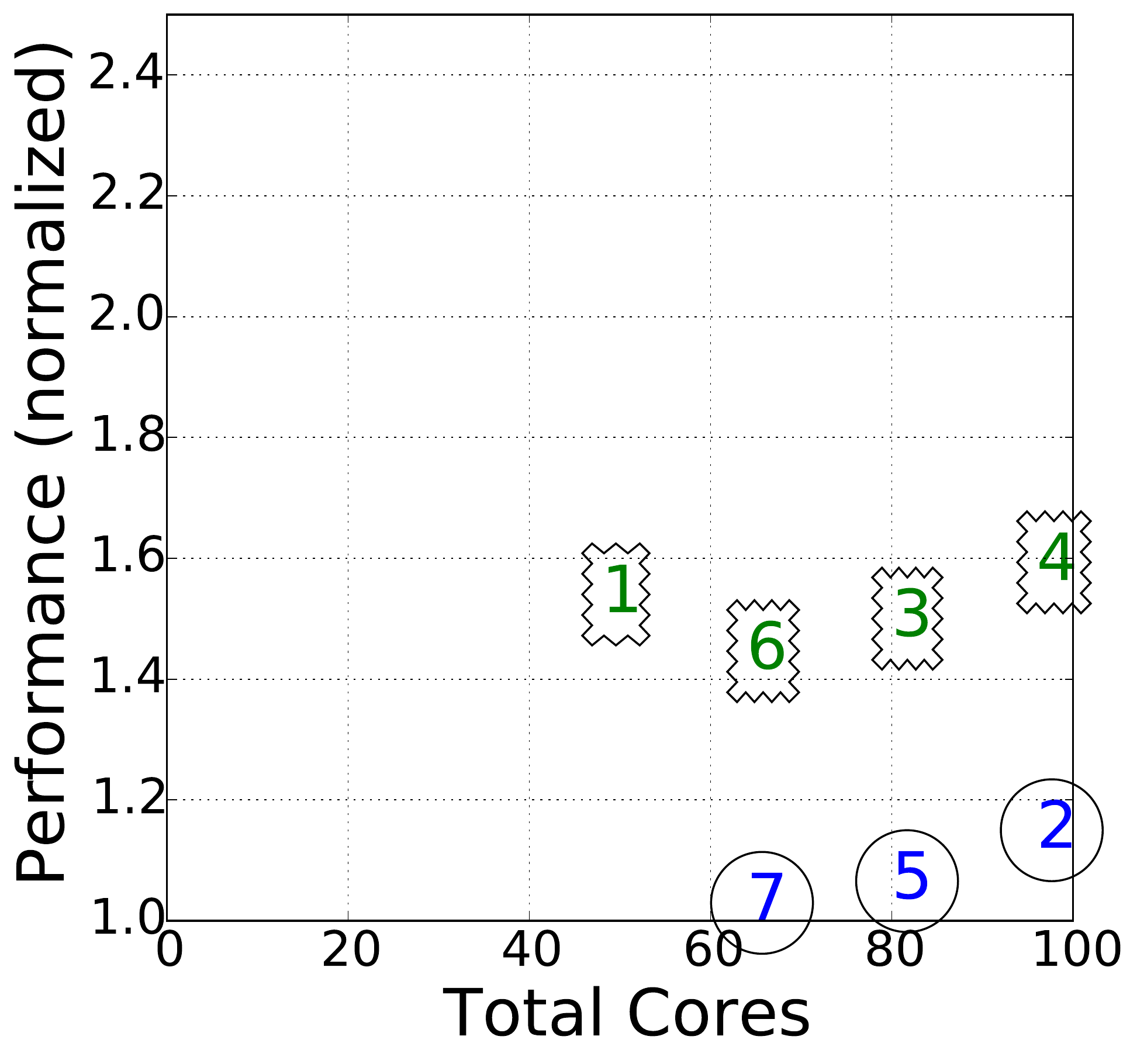}
 }
 \centering
 \caption{\small{\textbf{Minimizing the running cost for Naive-Bayes on Spark.} This is a memory-intensive workload.  \scout does not even try the \emph{c4} family due to its small memory per core.}}
 \label{fig:compare_4}
 \vspace*{-4mm}
\end{figure}

\noindent\textbf{Reliable exploration is difficult and generates high search cost.}
In Figure~\ref{fig:compare_2},~\ref{fig:compare_3},~\ref{fig:compare_1},~\ref{fig:compare_4}, we observe that the search path generated by \emph{CherryPick} involves more distinct VM types due to the need to explore the performance model. For example, in Figure~\ref{fig:compare_3}, CherryPick visits each instance family at once in all examples while \scout skips some specific families.
This is because \scout builds the performance model from historical data. Hence, it requires only little (or no) exploration. This phenomenon, exploration-exploitation dilemma, is studied extensively in Machine learning~\cite{kaelbling1996reinforcement}. The cold-start issue (as described in Section~\ref{sec:motivation} arises partly because of the requirement to explore the configuration space since \scout learns the performance behavior from historical data from workloads (previously explored) can sidestep the need to explore the search space.

\noindent\textbf{Fragility of CherryPick.}
As explained in Section~\ref{sec:motivation}, CherryPick is fragile
because it is sensitive to its parameters and the starting points.
In the four examples, \emph{CherryPick} starts from the same three configurations; however, the results are very different.
In \myfigure{\ref{fig:compare_3}}, \emph{CherryPick} fails to characterize the search space, which results in long search path (and high search cost).
While in \myfigure{\ref{fig:compare_4}},
\emph{CherryPick} stops too early and only finds a local minima (the $c4$ family).
These two examples show that \emph{CherryPick} is fragile and therefore, its search performance is not stable.

\noindent{\textbf{Scout identifies resource requirements}}:
When resource requirements can be articulated, a search process is more likely to find cloud configurations effectively and efficiently. In \myfigure{\ref{fig:compare_1}},
the \emph{PageRank} workload runs faster on a larger cluster (higher core counts) and higher-frequency CPUs. The \emph{r4} family, with larger memory but slower CPU speed, does not seem to be the best choice, hence avoided by \scout and instead prefers \emph{c4} and \emph{m4} family.
This tendency is more clear in the other cases as well (Figure~\ref{fig:compare_2}, ~\ref{fig:compare_3}, and~\ref{fig:compare_4}).

\noindent\textbf{Scout captures the complex cost model.}
In a real-world setting, practitioners can choose either a smaller cluster built using more powerful instances or choose large cluster built using smaller or less powerful instances (a scale-out and a scale-out configuration). The performance model used by \scout can infer the size of the cluster of the best cloud configuration.
In \myfigure{\ref{fig:compare_3}}, \scout chooses to run \emph{TeraSort} on a smaller cluster to save cost. On the contrary, in Figure~\ref{fig:compare_4}, \scout selects a larger cluster for efficiently running the \emph{Naive-Bayes} workload while achieving lower cost. These two examples show that \scout captures the complex relationship between the resource metrics and the running cost.

\noindent{\textbf{Summary}}:
The main difference between \emph{CherryPick} and \scout lies how the method explores the space of possible cloud configuration options. We can see that \emph{CherryPick} has to explore more cloud configuration options and hence have higher search cost (longer search path) while \scout searches within a relatively restricted region. This feature of \scout can be attributed to its performance model, which learns from the historical data. This also goes to show that encoding scheme, which uses low-level performance metrics, is successful in transferring knowledge from one workload to another.

\vspace{-0.4cm}
\section{Why \scout works better?}
\vspace{-0.4cm}

\scout relies on quality routing policy to deliver good solutions.
We find \scout effective because
it knows when to stop searching and
converges to better solutions.

\noindent{\textbf{Scout knows when to stop.}}
When an optimizer can stop as soon as it finds the optimal solution (or near-optimal solutions),
it can avoid unnecessary search efforts. \myfigure{\ref{fig:single_startingpoint}} shows that \scout requires a fewer number of steps if the starting point is already the optimal configuration.

\noindent{\textbf{Convergence speed.}}
The speed of convergence of a search-based method is dependent how it selects the next cloud configuration to measure. An ideal search-based method will always find the next cloud configuration, which is better than the cloud configurations sampled previously. \textit{Converge speed} can be defined as the average difference between the performance score (execution time or deployment cost) of the previous measurement (i$^{th}$ step) and the current measurement (i+1$^th$ step). A positive number would indicate that the current cloud configuration is better than the previous measurement (for both deployment cost and execution time, lower is better). \myfigure{\ref{fig:search_convergence}} compares the convergence speed of CherryPick and \scout. \myfigure{\ref{fig:search_convergence}} indicates that \scout overall finds cloud configurations 50\% (median) better execution time than the current cloud configuration, whereas CherryPick overall moves to cloud configuration which is 25\% worse than the current best configuration. Similar behavior is seen for deployment cost. This is evidence to show that \scout uses the historical data to find the promising region in the search space and exploits that space effectively.

\vspace{-0.4cm}
\section{Discussion}
\label{sec:discussion}
\vspace{-0.4cm}

%\subsection{Into \scout}
%\vspace{-0.15cm}
%\label{sec:into_scout}

\begin{figure}[t]
 \centering
 \subfigure[Search Performance]{
 \label{fig:single_cost_tuning_threshold_performance}
 \includegraphics[width=0.2\textwidth]{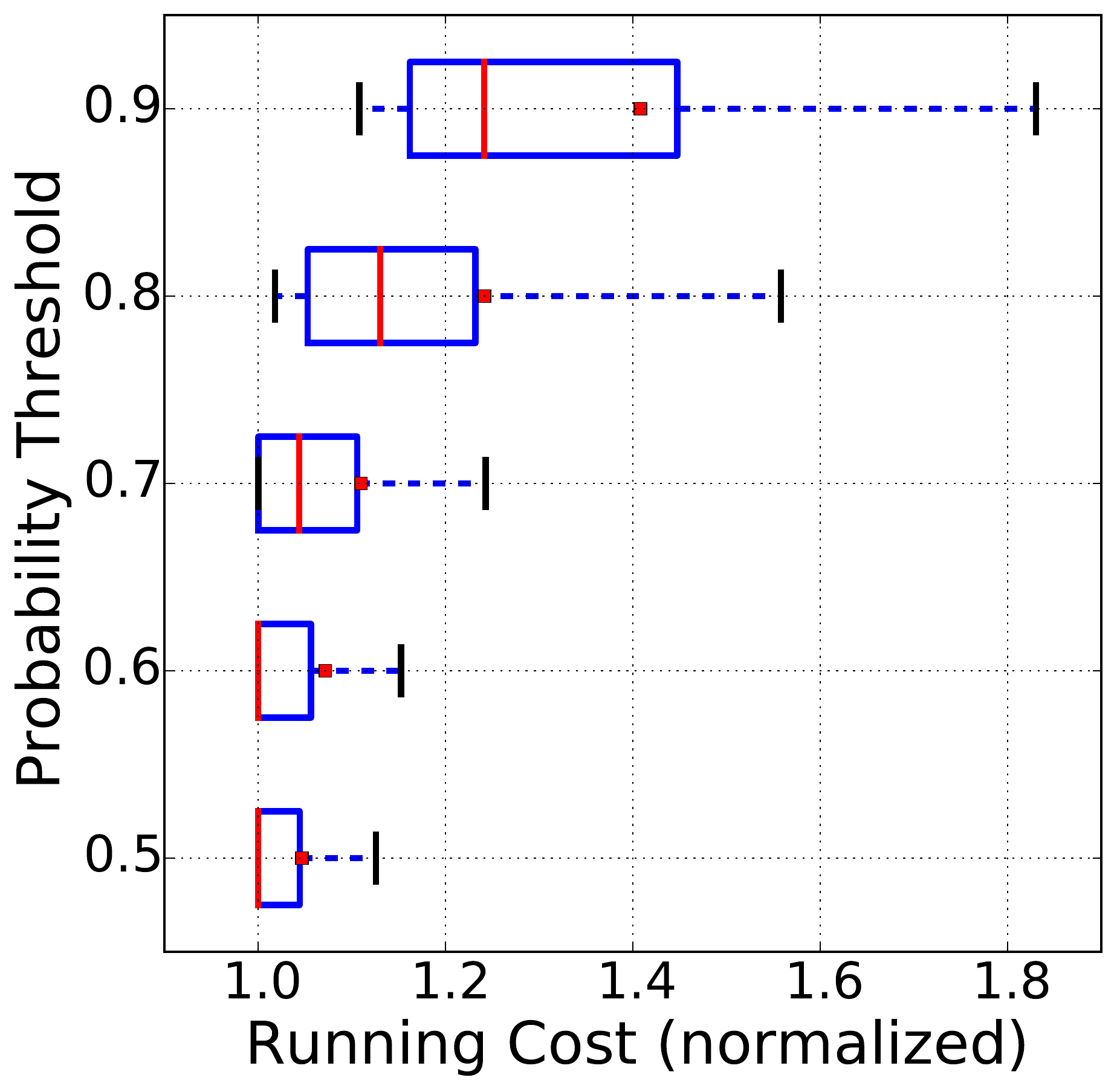}
 }
 \subfigure[Search Cost]{
 \label{fig:single_cost_tuning_threshold_steps}
 \includegraphics[width=0.2\textwidth]{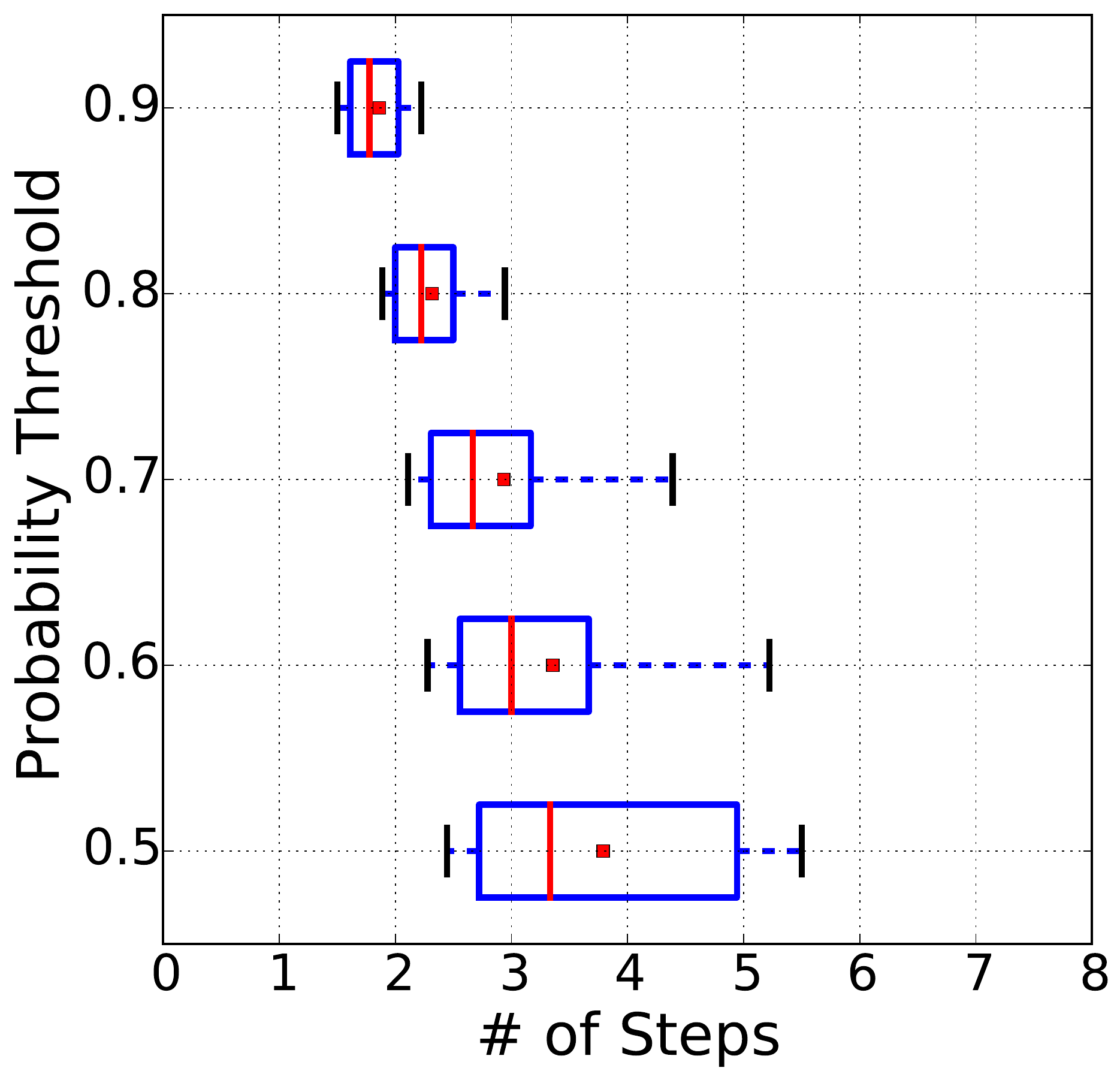}
 }
 \centering
 \caption{\small{\textbf{Tuning the probability threshold.} A smaller threshold generates longer search path but ensures better search performance.}}
 \label{fig:single_probability_threshold}
 \vspace*{-4mm}
\end{figure}

\noindent{\textbf{Tuning Searching Performance.}}
\scout uses ``probability threshold'' and ``misprediction tolerance'' as stopping criteria.
We examine how they affect the search performance of \scout.\\
\noindent{\textit{Probability Threshold}}:
\scout chooses the next cloud configuration to evaluate based on the probability of improvement and stops when the probability is lower than the probability threshold $\alpha$.
\myfigure{\ref{fig:single_time_overall}} shows that a higher probability threshold has pessimistic and terminates the search process prematurely, hence, shorter search path (as shown in Figure~\ref{fig:single_cost_tuning_threshold_steps}) and unstable search results (as shown in Figure~\ref{fig:single_cost_tuning_threshold_performance}). 
The probability threshold presents a trade-off between
the search performance and search cost.
Please note, the right threshold must consider the reliability curve of classification methods~\cite{niculescu2005predicting}.

\begin{figure}
 \centering
 \subfigure[Search Performance]{
 \label{fig:single_cost_tuning_tolerance_performance}
 \includegraphics[width=0.2\textwidth]{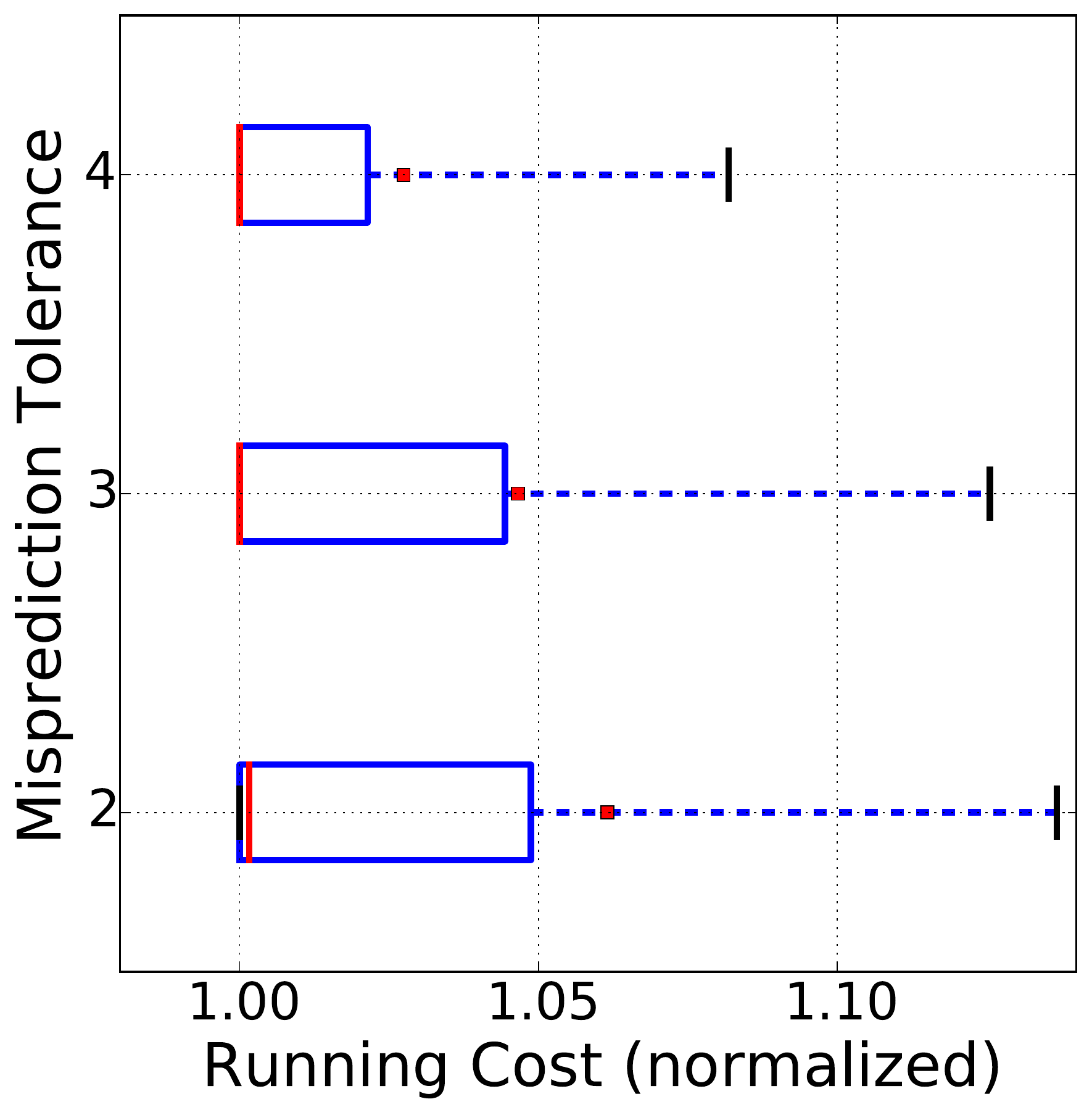}
 }
 \subfigure[Search Cost]{
 \label{fig:single_cost_tuning_tolerance_steps}
 \includegraphics[width=0.2\textwidth]{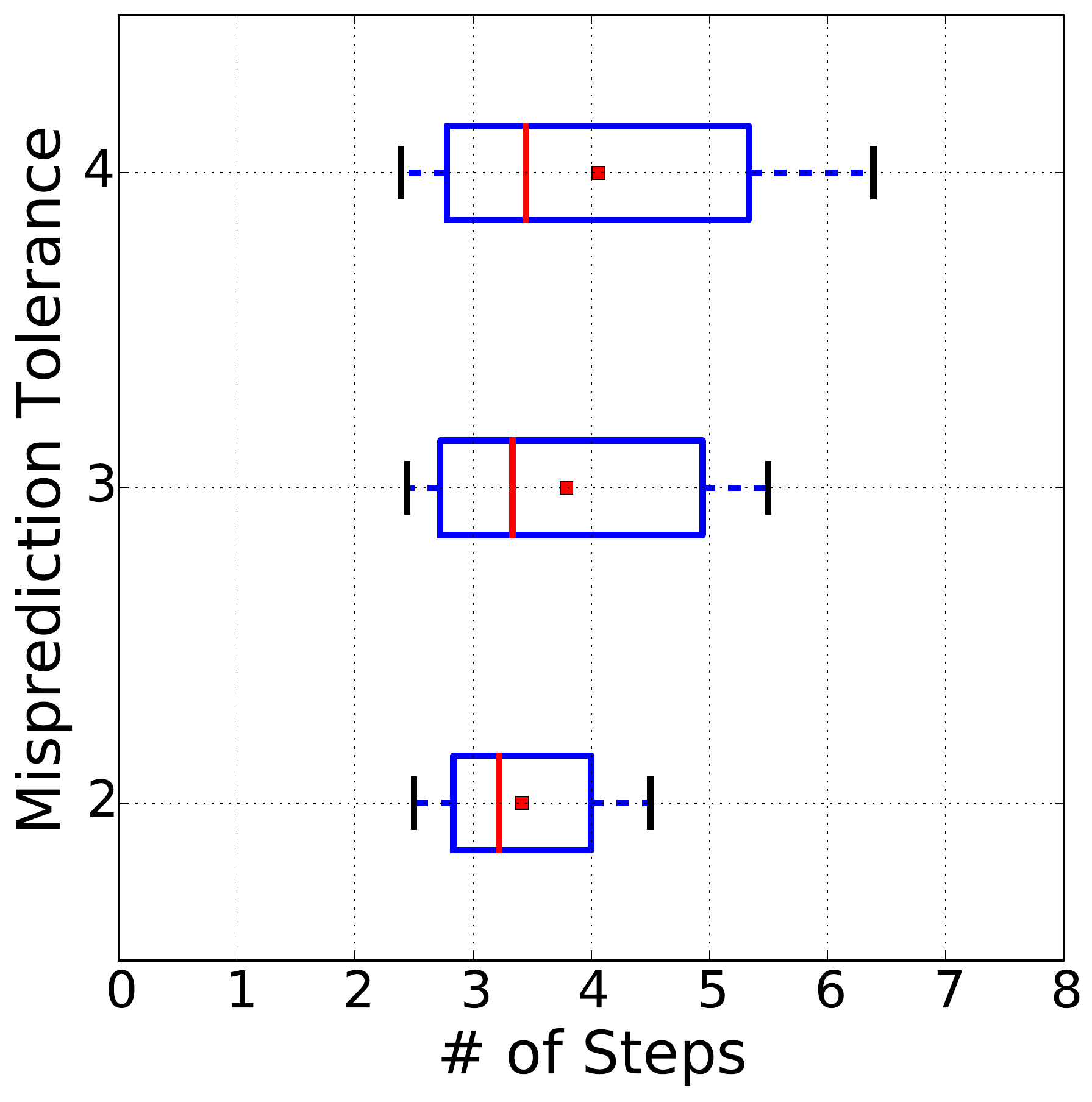}
 }
 \centering
 \caption{\small{\textbf{Tuning the misprediction tolerance.} A higher tolerance to mispredictions generates higher search cost.}}
 \label{fig:single_misprediction_tolerance}
 \vspace*{-4mm}
\end{figure}

\noindent{\textit{Misprediction Tolerance.}}
\scout terminates the search process if the selected configurations do not improve the current best choice (considered as a misprediction).
\scout uses an up limit and maintains a counter of mispredictions.
A larger limit tolerates more mispredictions but yields better search performance due to more chances.
A proper limit should consider both
the size of search space and the accuracy of prediction.
In \myfigure{\ref{fig:single_misprediction_tolerance}},
we show that a higher tolerance level leads to better search performance but higher search cost.
This trade-off is similar to the probability threshold.

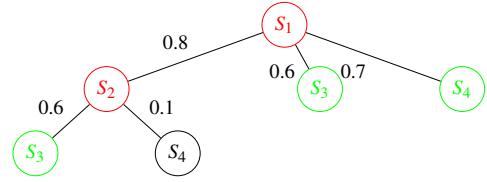
\begin{figure}
\small{
\resizebox{0.8\linewidth}{!}{%
 \begin{tikzpicture}[every tree node/.style={draw,circle},
   level distance=1cm,sibling distance=1.5cm, 
   edge from parent path={(\tikzparentnode) -- (\tikzchildnode)}]
 \Tree [.\node[red] {$S_1$};
         \edge node[auto=right] {0.8}; 
     [.\node[red] {$S_2$};
       \edge node[auto=right] {0.6};  
       [.\node[green] {$S_3$};]
       \edge node[auto=left] {0.1};
       [.$S_4$ ]
     ] \edge node[auto=right] {0.6}; 
     [.\node[green] {$S_3$};
     ] \edge node[auto=right] {0.7}; 
     [.\node[green] {$S_4$};
     ] ]
 \end{tikzpicture}
}}
     \centering
     \caption{\small{\textbf{An alternative search strategy.} For each observation ($S_{1}$ and $S_{2}$), \scout creates prediction probability for all configurations that are not evaluated yet. These predictions can be aggregated for selecting the next step.}}
     \label{fig:search_strategies}
\end{figure}

\noindent{\textbf{Alternative search strategies.}}
During search, \scout generates $P_{ij}$ for each new observation.
Our current search strategy only uses information from the latest observation.
\scout stores historical observations and therefore,
the next search step can be determined using several past observations.
In \myfigure{\ref{fig:search_strategies}}, we illustrate a way to incorporate other observations.
Given two observations on $S_1$ and $S_2$ and two unevaluated configuration $S_3$ and $S_4$,
\scout generates prediction probability, for example, 
we can generate prediction probability $P_{13}$, $P_{14}$ and $P_{23}$ and $P_{24}$.
Instead of choose $P_{23}$ after the second step,
\scout should choose $P_{14}$ when $S_2$ is much worse than $S_1$ (due to mispredictions).
This strategy is more likely to avoid bad choices.
On the other hand, 
\scout purely relies on offline performance modeling.
Another alternative is to update the prediction model upon new observations.
For unseen workloads, this update enables \scout to improve prediction accuracy.
However, the downside is the cost of retraining the model.
An online learning method might help reduce the retraining cost.
The two possible alternatives remain as future work.

%\vspace{-0.3cm}
%\subsection{Implications}
%\vspace{-0.15cm}

\begin{figure}[t]
 \includegraphics[width=.4\textwidth]{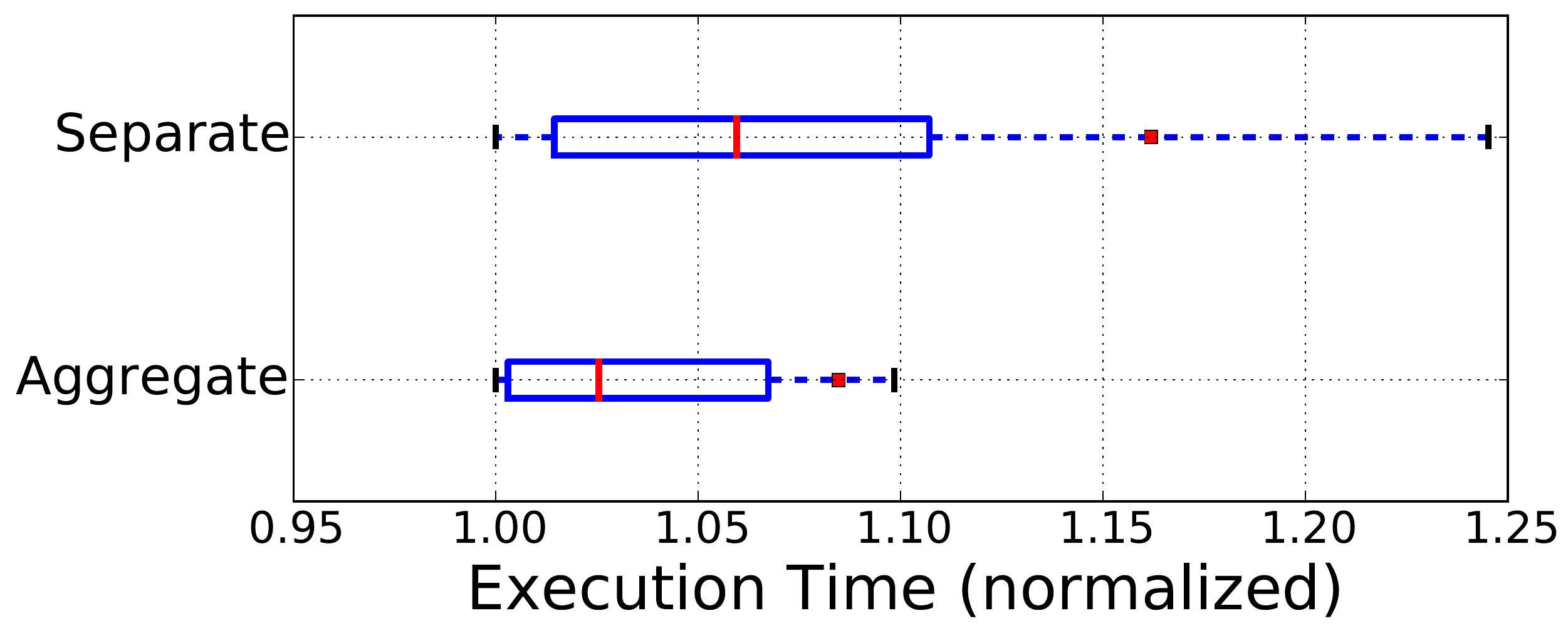}
 \centering
 \caption{\textbf{Universal performance models.}
 Prediction models built from multiple systems are more effective for \scout.}
 \label{fig:prediction_accuracy_comparison}
 \vspace*{-4mm}
\end{figure}
\noindent{\textbf{Universal Prediction Models.}}
% The performance models are built to find the best cloud configuration for a certain workload. These performance models generalize the information learned from the evaluated configurations. This helps the model to be accurate while predicting new cloud configurations. 
In prior work, the performance model needs to be retrained for every optimization process, which leads to wasted effort. There is a need for a modeling strategy, which becomes more accurate with experience. 
% This experienced model is useful in our setting, since measuring a new cloud configuration can be resource intensive. 
Transfer learning can be beneficial in our setting, where the performance model can predict from a new workload from data learned from performance data gathered while optimizing other workloads~\cite{pan2010survey}. \scout tries to learn from other performance data so that all the experience from the past optimization process is not lost.
Figure~\ref{fig:prediction_accuracy_comparison} shows how the performance model learned from more data (from different workloads) can generalize better than the performance model training from a single application. In the figure, the horizontal axis represents the execution time of the workload, and the vertical axis shows the two version of \scout. Separate refers to the \scout which is trained with performance data from just Hadoop workloads, whereas Aggregate refers to \scout trained on Hadoop as well as Spark workloads. We can see that Aggregate can find cloud configurations with better performance (lower execution time). This goes to show that in fact, more data is useful while training \scout. \textit{Overall, the prediction model used in \scout is a universal and can learn from any workload.}

\noindent{\textbf{Time-cost trade-off.}}
Users are willing to wait for a longer execution if a cloud configuration saves more (less cost).
For example, users are willing to trade a 20\% slow down in execution time for a 50\% decrease in running cost.
This is similar to the energy-time trade-off in high-performance computing~\cite{Freeh2007}.
\scout can support this scenario.
In our design, we define prediction classes based on the normalized performance of a single performance measure, \ie{time or cost}.
We instead define the classes based on the normalized performance of the product of time and cost
In our previous work, we show that how to support time-cost trade-off in a search-based method~\cite{Hsu2017}.
We plan to support this feature in \scout.

\vspace{-0.4cm}
\section{Related Work}
\label{sec:relatedwork}
\vspace{-0.2cm}

We already presented the most relevant work, \emph{CherryPick}~\cite{Alipourfard2017} and \emph{PARIS}~\cite{Yadwadkar2017}, in Section~\ref{sec:motivation}.
This section describes other optimization methods related to our work.

\noindent{\textbf{Cloud Deployment}}:
Cloud providers recommend the choice of VM types~\cite{aws, google_rightsizing}.
However, it is too coarse grain and does not
apply to many workloads because
resource requirement is often opaque~\cite{Yadwadkar2017}. \emph{Ernest} exploits the internal structure of the workload
to predict execution time of a workload~\cite{Venkataraman2016}.
\emph{Ernest} only needs smaller input for prediction.
This significantly reduces measurement cost.
However, \emph{Ernest} is not scalable because
the prediction model is specific to a VM type.

\noindent{\textbf{Parameter Tuning}}:
System and software performance is highly affected by configurations.
StarFish is an auto-tuning system for Hadoop applications~\cite{herodotou2011starfish}.
\emph{BestConfig} proposes the Divide and Diverge Sampling strategy along with the Recursive Bound and Search method
for turning software parameters~\cite{zhu2017bestconfig}.
Similar framework is also proposed to automate tuning system performance
of stream-processing systems~\cite{bilal2017towards}.
BOAT is a structured Bayesian Optimization-based framework for automatically tuning system performance
~\cite{Dalibard2017} which leverages contextual information.
Sampling techniques focus on reducing sampling cost while building accurate models to optimize software systems \cite{nair2018finding, oh2017finding, Nair2017}.
Parameter tuning is also an critical in machine learning~\cite{Dewancker2015,shahriari2016taking,Klein2017,golovin2017google}.

\noindent{\textbf{Sampling Methods}}:
Sampling techniques focus on reducing sampling cost while building accurate models to optimize software systems \cite{nair2018finding, oh2017finding, Nair2017}.
The above methods reduce the search cost by a significant degree.
However, they focus on performance tuning for the same workload (or application) on the same type of machine. It is not clear how to leverage their approaches to support different
machine configurations in cloud computing. We, instead, find the best machine configuration for a given workload.

%\noindent{\textbf{Parameter Tuning}}:
%This is similar to configuration tuning.
%Many machine learning models require proper tuning
%of their parameters for improving prediction.
%Bayesian Optimization has been successfully applied
%to hyper-parameter tuning and used as a drop-in replacement to grid %search~\cite{Dewancker2015,shahriari2016taking,Klein2017}.
%\emph{Google Vizier} is a cloud service for black-box optimization.
%It uses Bayesian Optimization to tune the parameters
%for machine learning services.

\section{Threats to Validity}
As with any empirical study, biases can affect the final results. Therefore, any conclusions made from this work must
be considered with the following issues in mind:
\vspace{-0.5em}
\begin{enumerate}[leftmargin=*]
    \setlength\itemsep{-0.4em}
    \item \textit{Sampling bias} threatens any classification experiment; i.e., what matters there may not be true here. For example, our dataset is collected from our extensive experimentation based on our knowledge of the application and workloads. Also even though we use 107 different workload and 30 different applications they are all from three big data analytics systems.  Besides, they are JVM based systems. It is not clear whether our method applies to non-JVM workloads, which requires further investigation.
    \item  \textit{Learner bias:} For building the performance models, we elected to use ExtraTrees because it is the state-of-the-art machine learning method for high-dimensional problems (as in our modeling requirement of handling low-level performance data).  
    %Classification is a large and active field, and any single study can only use a small subset of the known classification algorithms.
    \item \textit{Evaluation bias}: This paper uses three measure of performance, execution time, running cost, and search cost (in terms of search steps). There could be other ways to measure the performance such as resource requirements, the actual cost of evaluation, etc.
\end{enumerate}
\vspace{-0.5em}

\vspace{-0.4cm}
\section{Conclusion}
\label{sec:conclusion}
\vspace{-0.2cm}

Today most applications are hosted in the cloud.
It is essential to maximize the performance of an application while keeping the deployment cost down.
Machine learning and sampling techniques have been previously proposed to
build models to predict the performance of cloud configurations.
However, the techniques proposed in prior work are either expensive to train or are unreliable---if trained on sparse samples. 

Our method, \scout, is different to the previously proposed directions and promotes learning from previous experience---optimization process.
We advocate using historical data to identify regions on the configuration space, which might contain the best cloud configuration. To use the historical data, we propose a new modeling scheme which use low-level metrics along with pair-wise modeling technique to transfer knowledge from one optimization process to the other.

In our experience, performance data is hard to find. A lack of performance data discourages the advances in system performance research. We believe that we will see advances in performance optimization by sharing performance data.
Our large-scale performance dataset is available at \url{https://github.com/oxhead/scout}.
%Also, please consider sharing your data by submitted a pull request to this repository.

%\input{acknowledgments.tex}

% \section{Availability}
% It's great when this section says that MyWonderfulApp is free software, 
% available via anonymous FTP from

{\footnotesize \bibliographystyle{acm}
%\bibliography{reference}}

%\theendnotes

\end{document}